\documentclass[particles,article,submit,moreauthors,pdftex]{Definitions/mdpi}

\usepackage{hyperref}
\usepackage{amsmath}
\usepackage{amsfonts}
\usepackage{amssymb}
\usepackage{graphicx}
\usepackage[utf8]{inputenc}
\usepackage[T1]{fontenc}
\usepackage[titletoc]{appendix}

\usepackage{color}

\firstpage{1}

\Title{Simulating Binary Neutron Stars with Hybrid Equation of States: 
Gravitational Waves, Electromagnetic Signatures, and 
Challenges for Numerical Relativity}

\Author{Henrique Gieg$^{1,*}$, Tim Dietrich$^{2}$, Maximiliano Ujevic$^{1}$}

\address{${}^1$ \quad Centro de Ci\^encias Naturais e Humanas,  
Universidade Federal do ABC, 09210-170, Santo Andr\'e, S\~ao Paulo, Brazil \\
${}^2$ \quad Nikhef, Science Park, 1098 XG Amsterdam, The Netherlands }

\corres{Correspondence: henrique@gieg.com.br}

\abstract{The gravitational wave and electromagnetic signatures 
connected to the merger of two neutron stars allow us to test the nature 
of matter at supranuclear densities. Since the Equation of State 
governing the interior of neutron stars is only loosely constrained, 
there is even the possibility that strange quark matter exists inside 
the core of neutron stars. We investigate how strange quark matter cores 
affect the binary neutron star coalescence by performing numerical 
relativity simulations. Interestingly, the strong phase transition can 
cause a reduction of the convergence order of the numerical schemes to 
first order if the numerical resolution is not high enough. Therefore, 
an additional challenge is added in producing high-quality gravitational 
wave templates for Equation of States with a strong phase transition.
Focusing on one particular configuration of an equal mass configuration 
consistent with GW170817, we compute and discuss the associated 
gravitational wave signal and some of the electromagnetic counterparts 
connected to the merger of the two stars. We find that existing waveform 
approximants employed for the analysis of GW170817 allow describing this 
kind of systems within the numerical uncertainties, which, however, are 
several times larger than for pure hadronic Equation of States, which 
means that even higher resolutions have been employed for an accurate 
gravitational wave model comparison. We also show that for the chosen 
Equation of State, quasi-universal relations describing the 
gravitational wave emission after the moment of merger seem to hold and 
that the electromagnetic signatures connected to our chosen setup would 
not be bright enough to explain the kilonova associated to GW170817.}


\begin{document}

\section{Introduction}

The observation of gravitational waves (GWs), GW170817, and electromagnetic (EM) signatures, 
AT2017gfo and GRB170817A, originating from the same astrophysical source have initiated 
a new era of multi-messenger astronomy~\cite{TheLIGOScientific:2017qsa,GBM:2017lvd,
Monitor:2017mdv}. 

Because of their large compactness, neutron stars (NSs) contain matter at supranuclear 
densities, which makes them a perfect laboratory to determine matter under 
extreme conditions governing the NS interior. 
Today's Equation of State (EoS) constraints inferred from GW170817 arise 
from the analysis of the GW signal, e.g.,~\cite{TheLIGOScientific:2017qsa,Dai:2018dca,
De:2018uhw,Abbott:2018wiz,Abbott:2018exr,LIGOScientific:2018mvr} or 
from a combined GW and EM analysis, 
e.g,~\cite{Radice:2017lry,Bauswein:2017vtn,Coughlin:2018miv,
Radice:2018ozg,Coughlin:2018fis}.  
In addition, one can also perform statistical analyses of a large 
set of possible EoSs incorporating information from 
nuclear physics theory, e.g.,~\cite{Annala:2017llu,Most:2018hfd}. 

The current analysis of GW170817 disfavors NSs 
with too large radii and tidal deformabilities, 
the exact state of the supranuclear matter inside the NS core is still unknown. 
While recent first-principle studies disfavor the presence of strange quark matter (SQM) inside NSs 
with masses $\lesssim 1.4M_\odot$, Ref.~\cite{Annala:2019puf}, it is not clear 
if more massive NSs could have SQM cores or if the merger could undergo a strong phase 
transition~\cite{Most:2018eaw,Bauswein:2018bma}. 

In this article, we study possible effects of SQM during the binary coalescence, for this purpose, we solve 
the equations of general relativity combined with the equations of general relativistic 
hydrodynamics (GRHD) with the help of full 3+1 -dimensional 
(3 dimensions in space and 1 dimension in time) numerical relativity (NR) 
simulations. Such simulations allow a characterization of the GW and EM radiation. 
Over the last years, the NR community has made significant 
progress in simulating neutron star spacetimes on many 
fronts, e.g.,~extending the simulation's accuracy, incorporating 
microphysical models in the simulations, and exploring different 
regions of the binary neutron star (BNS) parameter space~\cite{Radice:2013hxh,
Hotokezaka:2015xka,Dietrich:2017aum,Kiuchi:2017pte, Dietrich:2018phi,
Rezzolla:2011da,Neilsen:2014hha,
Sekiguchi:2015dma,Palenzuela:2015dqa,Foucart:2017mbt,
Palenzuela:2015dqa,Ruiz:2017due,Ciolfi:2017uak,
Kiuchi:2017zzg,Radice:2017zta,Shibata:2017jyf}. 
We will show that standard techniques are capable of 
dealing with NSs containing hadronic and SQM. 
Up to our knowledge, most previous simulations with EoSs allowing 
strong phase transitions have studied the imprint of 
a SQM phase after the merger of the two 
stars~\cite{Most:2018eaw,Bauswein:2018bma}, in addition, also 
the phenomenological implications of the two-families scenario
(in which hadronic stars and pure strange quark stars
coexist) has been recently analysed~\cite{Pietri:2019eb}, 
but no numerical relativity inspiral simulation with stars 
containing SQM has been performed.
Thus, the effect of employing a hybrid EoS, in which
hadrons convert into strange quarks through a
thermodynamical phase transition, on the inspiral, 
has to be investigated in more detail.
We find that a strong phase transition impairs the 
NR inspiral evolution and that the convergence order of the simulations 
can drop to first order if the employed resolutions are not sufficient. 
While it might be possible to deal with phase transitions incorporating 
more sophisticated numerical techniques or higher resolutions,
our work is (up to our knowledge) the first to point out 
this arising complication. 

The structure of the article is as follows. 
First, we discuss how we construct a hybrid star EoS using the tdBag model, 
we continue by describing the employed setup and numerical techniques. 
Then, we present a first qualitative description of the binary 
coalescence and continue by discussing the GW emission during the 
inspiral and postmerger phase, as well as, the related ejecta outflow 
and expected EM counterparts. 
Unless otherwise stated, we employ geometric units, i.e., 
$G=c=1$. Physical units are sometimes given for better intuition. 

\section{A new hybrid Star Equation of State}

\subsection{tdBag model for strange quark matter}

tdBag~\cite{Farhi:1984qu} is a bag-like model inspired by the standard 
MIT bag model~\cite{Chodos:1974je,DeTar:1979vb}\footnote{For convenience 
we will use natural units in this section setting $\hbar = c = 1$.}. It 
incorporates features expected for cold strange quark matter composed of 
up, down, and strange quarks in free Fermi seas resulting from the 
deconfinement, i.e., the breaking of hadronic matter into its 
constituents once baryon densities compatible to the nuclear density 
$n_{0} \sim 0.16$ fm$^{-3}$ are reached. tdBag accounts for perturbative 
effects of strong interaction between quarks~\cite{Alford:2004pf} and 
the effects of color superconductivity at low 
temperatures~\cite{Alford:2007xm}. There are many possible color 
superconducting phases, in this work we assume that quarks are in the 
color-flavor-locked phase (CFL) \cite{Lugones:2002va,Alford:2007xm}, in 
which the three quark flavors are symmetric (equal baryon density of 
up, down and strange) and exhibit superfluidity.

In the massless up and down quarks approximation $m_{u} = m_{d} = 0$, 
the tdBag model is defined by the grand potential per volume ($\Omega$) as
 \begin{equation}
 \Omega = -\frac{3}{4\pi^2}a_{4}\mu^4 + \frac{3}{4\pi^2}a_{2}\mu^2 + B 
+\Omega_{e},\label{eq:potential}
 \end{equation}
where $a_{4}$, $a_{2}$, and $B$ are free parameters and $\mu 
\equiv (\mu_{u} + \mu_{d} + \mu_{s})/{3}$ is the quark chemical 
potential, $\mu_{i}$ refers to the chemical potential of the $i$-th 
flavor, and 
$\Omega_{e}$ is the electrons' contribution. Before proceeding to the 
construction of hybrid EoS containing a SQM core, 
we are going to briefly discuss the meaning of the 
parameters presented in Eq.~\eqref{eq:potential}:
  
1. The quartic coefficient $a_{4}$ can be interpreted as a
correction to the pressure of the free Fermi sea arising 
from quantum chromodynamics (QCD). 
Fraga et al.~\cite{Fraga:2001id} show that, to second order in the strong 
coupling constant, a reasonable value for this parameter is $a_{4} \sim 
0.7$. The values of $a_{4} = 0$ and $a_{4} =1$ correspond to maximum or 
none QCD interaction, respectively. 

2. The quadratic coefficient $a_{2}$ accounts for the contribution of the strange quark 
mass and color superconductivity. The value of this parameter is given by $a_{2} = 
m_{s}^2 -4\Delta^2$, where $m_{s} \sim 100$~MeV is the strange quark 
mass and $\Delta \sim 0 - 100$~MeV is the energy gap between free and 
Cooper pairs of quarks~\cite{Alford:2004pf,Lugones:2002va,Klahn:2015mfa}.

3. The bag constant $B$, introduced phenomenologically into the theory
to describe the deconfinement, is the energy density required to produce a 
``vacuumless'' volume in which quarks exist. 

4. The contribution of the electron potential ($\Omega_{e}$) for our 
purposes is negligible; cf.~Ref.~\cite{Pereira:2017rmp} and references 
therein. Nevertheless, if the SQM is not in the CFL 
phase~\cite{Lugones:2002va} there must be electrons within the quark 
phase to ensure (i) local charge neutrality and (ii) chemical 
equilibrium. These two conditions may be expressed respectively as
 \begin{eqnarray}
 \frac{2}{3}n_{u} - \frac{1}{3}n_{d} - \frac{1}{3}n_{s} = n_{e}, \\ 
 \mu_d = \mu_s = \mu_{u} + \mu_{e},
 \end{eqnarray}
where $n_{i}$ denotes the number density. For SQM in the CFL phase, 
the quark matter is symmetric, thus, local charge neutrality is satisfied 
without the presence of electrons. In both cases, CFL ($\Omega_e=0$) or 
non-CFL phase (neglecting $\Omega_e$), Eq.~\eqref{eq:potential} leads to an 
analytical EoS for SQM.

\subsection{Thermodynamics and parameter analysis}

In this section we constrain the parameter space $(a_{4},a_{2},B)$ 
presented in Eq.~\eqref{eq:potential} assuming the absolutely stable 
strange quark matter hypothesis \cite{Bodmer:1971we,Witten:1984rs} and thermodynamic 
properties. With the help of the grand potential $\Omega$, we found that 
the pressure $p$, the 
baryon number density $n_{b}$, and the energy density $\epsilon$ can be 
written as
 \begin{eqnarray}
 p &=& -\Omega = \frac{3}{4\pi^2}a_{4}\mu^4 - 
\frac{3}{4\pi^2}a_{2}\mu^2 - B, \label{eq:p} \\
 n_{b} &=& -\frac{1}{3}\frac{\partial \Omega}{\partial \mu} = 
\frac{1}{2\pi^2}(2a_{4}\mu^3 - a_{2}\mu),\label{eq:n} \\
 \epsilon &=& -p + 3\mu n_{b} = \frac{9}{4\pi^2}a_{4}\mu^4 - 
\frac{3}{4\pi^2}a_{2}\mu^2 + B. \label{eq:e}
 \end{eqnarray}

It is possible to obtain $p(\epsilon)$ from Eq.~\eqref{eq:p}, 
\eqref{eq:n} and \eqref{eq:e}, and the result is
 \begin{equation}
 p(\epsilon) = \frac{1}{3}(\epsilon -4B) -\frac{a_2^2}{12\pi^2a_4}\left( 
1+\sqrt{1+\frac{16\pi^2a_{4}}{a_{2}^2}(\epsilon - 
B)}\right), \label{eq:eosq}
 \end{equation}
which is the analytical EoS for SQM parametrized by $a_{4}$, 
$a_{2}$, and $B$.

As long as we are dealing with a phenomenological model, it is of 
fundamental importance to constrain the parameter space within a region 
in which it can produce EoSs for stellar matter. To justify our next 
choices, we make a digression on the absolutely stable strange quark 
matter hypothesis (i.e. no strange matter will decay into hadronic 
matter under a thermodynamic process), proposed by Bodmer \cite{Bodmer:1971we} 
and Witten \cite{Witten:1984rs}. They consider the ground state of matter to be 
three Fermi seas of deconfined quarks in equal proportions of up, down 
and strange flavors. The reason for this is that the energy per baryon 
($E/A$) in the ground state is smaller than that of the most stable 
pressureless nuclei, $^{56}$Fe with $E/A\sim 930$ MeV. Until now, no 
experimental observations could refute this hypothesis. Indeed, as 
deconfinement is expected to take place under extreme conditions such as 
the ones in the core of compact stars, many efforts have been made to 
uncover which observables in astrophysical phenomena could rule out or 
confirm the existence of SQM in stars.

In the case of a Hybrid Star (HyS), i.e., a star containing a SQM core surrounded by 
hadronic matter (HM), the deconfinement of HM into SQM must take place 
at a strictly positive pressure. 
Assuming that the deconfinement is a first order phase transition, i.e., 
the interface between the hadron and quark phases is characterized by a 
sharp discontinuity (no mixed phase is present). 
At the interface the Gibbs condition must be satisfied
\begin{equation}
 g_{q}(p) = g_{h}(p),
 \label{eq:gibbs-cond}
\end{equation}
where $g_{q}$, $g_{h}$ are the Gibbs free energy per baryon of 
the quarks and hadrons, respectively, with the generic $g$ given by
 \begin{equation}
 g = \frac{\epsilon + p}{n_{b}}=3\mu. \label{eq:g}
 \end{equation}
Eq.~\eqref{eq:gibbs-cond} represents chemical and mechanical 
equilibrium at the interface. Using Eqs.~\eqref{eq:g} and~\eqref{eq:p}, one can 
write $p$ as a function of the Gibbs free energy
 \begin{equation}
 p(g) = \frac{a_{4}}{108\pi^2}g^4 -\frac{9a_{2}}{108\pi^2}g^2 - B.
\label{eq:p(g)}
 \end{equation}

To enforce the phase transition pressure to be strictly positive 
($p>0$), the following relation must hold
 \begin{equation}
 B>\frac{g^2}{108\pi^2}(a_{4}g^2-9a_{2}).
 \end{equation}
According to the absolutely stable SQM hypothesis the minimum of $g$ 
is $g_{\rm min}(p=0) \approx 930$~MeV. This value of $g$ sets $B_{\rm 
min}$, for which hadronic matter deconfines in three quark flavors, such 
that
 \begin{equation} 
 B_{\rm min}=\frac{g_{\rm min}^2}{108\pi^2}(a_{4}g_{\rm min}^2 
-9a_{2}).\label{eq:Bmin1} 
 \end{equation}
This equation defines the region on the $B$, $a_{2}$, $a_{4}$ 
space above which a transition from HM to SQM happens.
On the other hand, the hadronic part of the HyS must not be in 
metastable equilibrium, i.e., one shall guarantee that it would not 
decay in 2-flavor quark matter. This condition can be ensured if
 \begin{equation}
 g_{2f}>g_{h},
 \end{equation}
where $g_{2f}$ is the Gibbs free energy per baryon for 
2-flavor quark matter. A second similar condition for $B$ can be 
found~\cite{Pereira:2017rmp} adapting Eq.~\eqref{eq:potential} to the $2f$ case in the 
massless approximation ($m_{u} = m_{d} = 0$), the result is
 \begin{eqnarray}
 B &>&\bar{B}_{\rm min}, \nonumber \\
 \bar{B}_{\rm min} &=& \frac{g_{\rm min}^2}{54\pi^2} \left[ 
\frac{4g_{\rm min}^2a_{4}}{(1 + 2^{1/3})^3}-3a_{2} \right].\label{eq:Bmin2}
 \end{eqnarray}

\begin{figure}[htpb]
\centering
\includegraphics[width=0.6\textwidth]{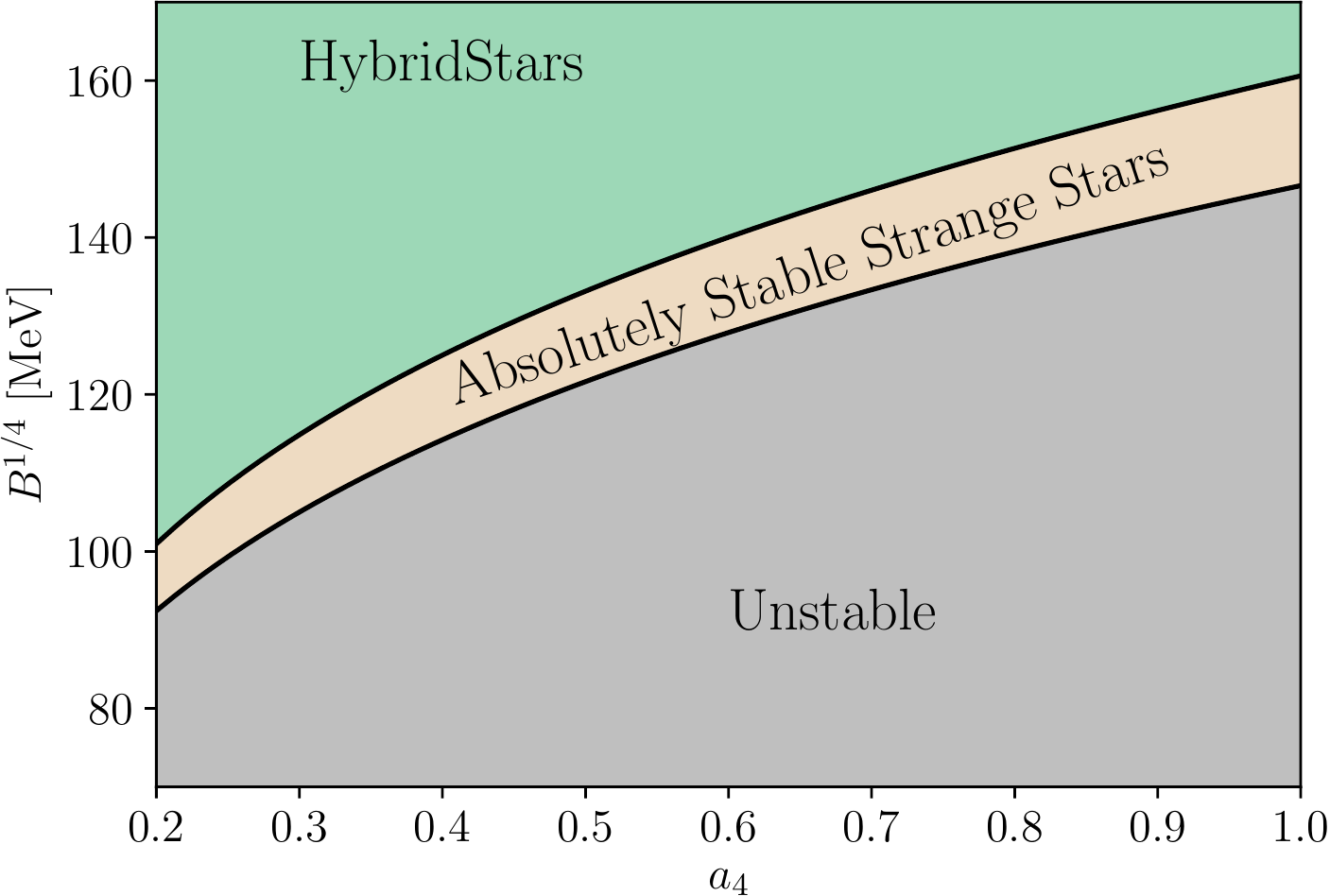}
\caption{Parameter space $(a_{4}, B)$ for fixed $a_{2} = 5000$ MeV$^2$. We 
identify three different regions. The upper region where HM 
$\rightarrow$ SQM transitions are possible and describes a HyS. The 
middle region where SQM is absolutely stable and we found strange stars. 
The lower region is prohibited due to the instability of SQM.}
\label{fig:par_space}
\end{figure}

Therefore, conditions~\eqref{eq:Bmin1} and \eqref{eq:Bmin2} constrain 
the parameters space $(a_{4}, a_{2}, B)$ in three regions (see Fig. 
\ref{fig:par_space} for an example),
 \begin{itemize}
 \item[(i)] $B<\bar{B}_{\rm min}$: not allowed region due to the fact 
that hadronic matter is more stable than quark matter, then we cannot 
use a SQM EoS.
 \item[(ii)] $\bar{B}_{\rm min} < B < B_{\rm min}$: absolutely stable 
strange stars (SS). There is no phase transition HM $\rightarrow$ SQM 
and the star is entirely made of 3-flavor quarks.
 \item[(iii)] $B>B_{\rm min}$: HyS. There exists a phase transition between HM 
 and SQM for some energy density $\epsilon$.
 \end{itemize}

The EoS for SS is given directly by Eq.~\eqref{eq:eosq}. 
To obtain the EoS for HyS we must solve Gibbs condition (\ref{eq:gibbs-cond}) 
with the help of a hadronic EoS to find the deconfinement pressure 
$p_{T}$ and the corresponding energy densities $\epsilon_{h}$ and 
$\epsilon_{q}$ for hadronic and quark matter, respectively. 
Assuming a first order phase transition, the HyS EoS is then 
constructed as a piecewise function with the prescription
 \begin{itemize}
 \item[(i)] $\epsilon \leqslant \epsilon_{h}$: HM EoS,
 \item[(ii)] $\epsilon_h < \epsilon < \epsilon_{q}$: $p = p_{T}$,
 \item[(iii)] $\epsilon \geqslant \epsilon_{q}$: SQM EoS.
 \end{itemize}

If deconfinement produces a sharp discontinuity between the phases 
or an extended region in which hadrons and quarks are mixed, 
is an open question that relies primarily on how global electric charge
neutrality is achieved by the star. This could happen in two scenarios:
(i) assuming local charge neutrality, i.e., that the electric charge
density is zero everywhere in the stellar interior or (ii) assuming that 
electric charge density is non-zero and distributed within the star
such that the total charge is zero. Despite the latter
being a relaxation of the first condition, it has profound implications
for the star's structure, e.g. the rise of Coulomb interactions
between quarks in different geometrical shapes (drops,
rods, and slabs) and the surrounding hadronic matter may lead
to a mixed phase of quarks and hadrons extending along 
the stellar interior \cite{Glendenning_book}. On the other hand,
if local charge neutrality is considered, quarks and hadrons
separate into two phases with a sharp discontinuity interface 
between them. The existence of a mixed phase is favored if the Coulomb
interactions were to lower the system's energy~\cite{Glendenning_book} in comparison
to maintain the interface by surface tension.
Recently, it had been shown~\cite{Lugones:2013ema} that it is energetically
favorable for SQM to sharp discontinuity interface if the quark matter
is described by the Nambu-Jona-Lasinio model~\cite{Nambu:1961tp}. 

\subsection{Building hybrid star EoS}

The recent observation of GWs from GW170817 established 
constraints on the NS EoS. The analysis of the data favored soft EoS 
instead of stiff EoS. For our analysis, we use for the HM phase the SLy 
EoS \cite{Douchin:2001sv}, which is currently in agreement with EoS constraints. 
Recalling that we assumed the deconfinement as a 
first order phase transition (at the sharp interface between quarks and 
hadrons) the Gibbs conditions \eqref{eq:gibbs-cond} must hold for both 
phases.

\begin{figure}[htpb]
\centering 
\includegraphics[width=0.49\textwidth]{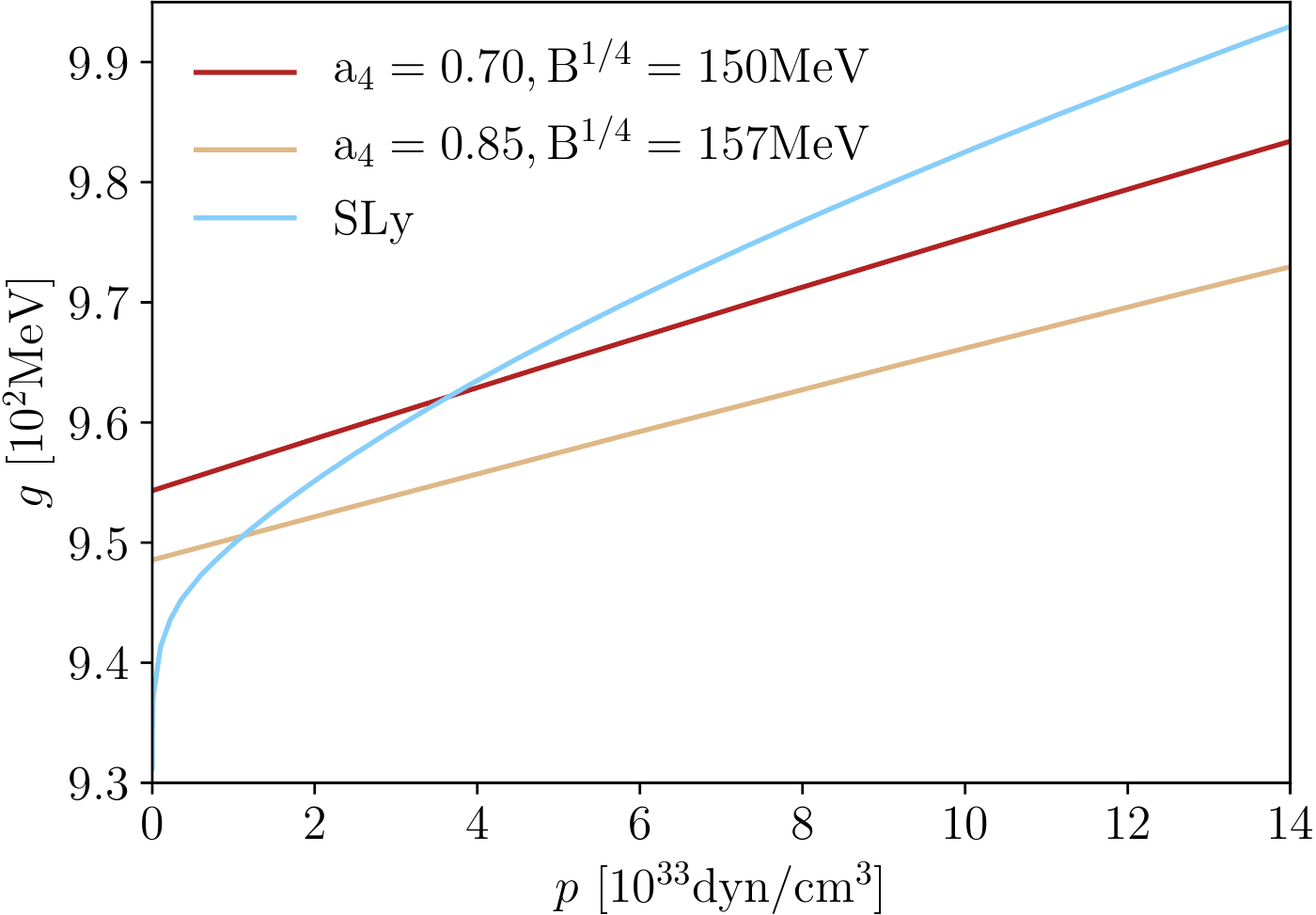} \quad
\includegraphics[width=0.48\textwidth]{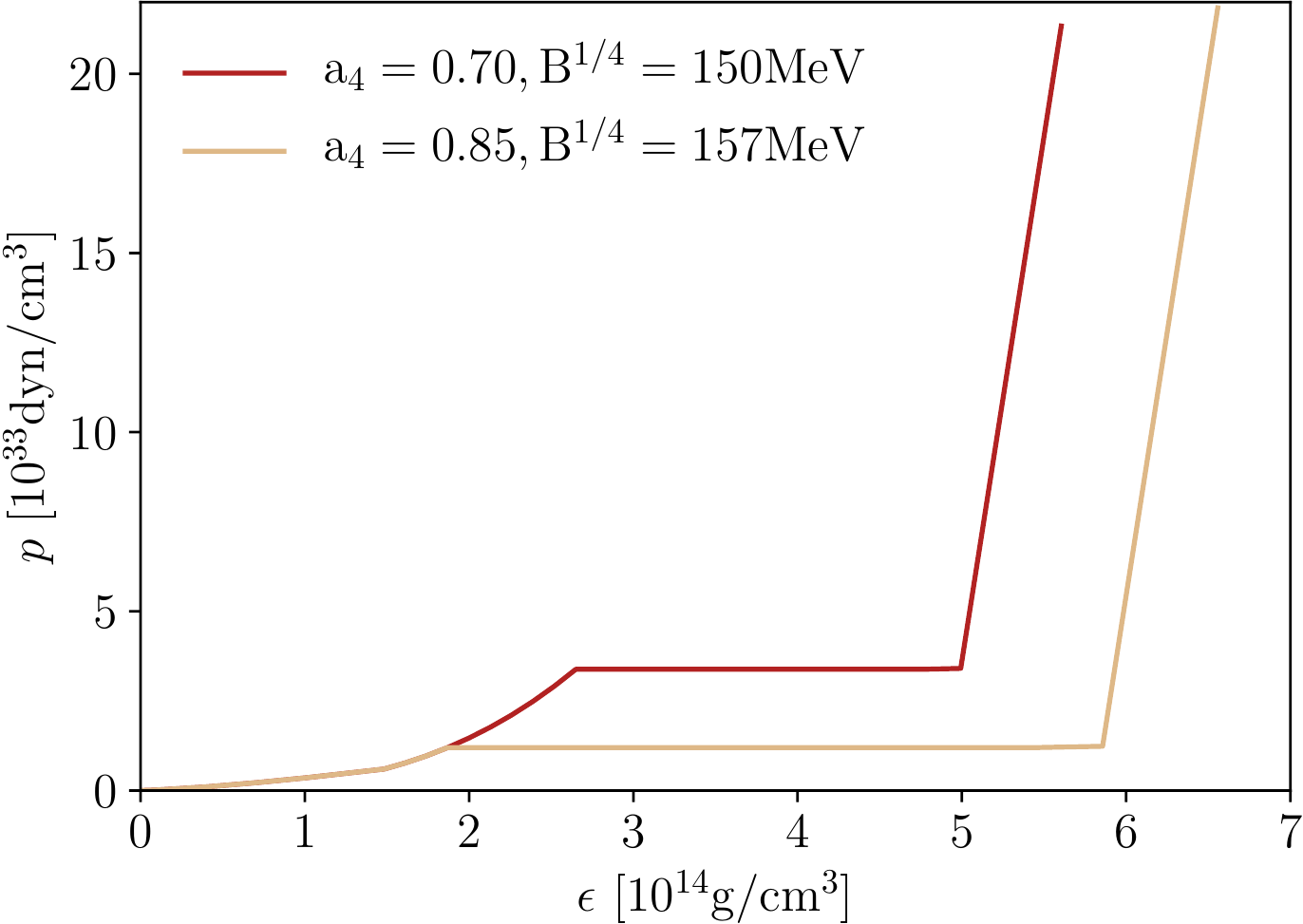}
\caption{Left panel: 
Gibbs free energy $g$ as function of pressure $p$. For the SQM 
EoS we set $a_{2} = 5000$ MeV$^2$. The point at which SLy and SQM curves 
intersect determines the transition pressure $p_{T}$ and the transition 
Gibbs free energy $g_{T}$. For pressures greater than $p_T$ the SQM free 
energy becomes lower than the SLy free energy, thus being energetically 
favorable. 
Right panel:
EoS for the HyS constructed by the junction of the SLy EoS and 
two parameter set for the SQM EoS. We assume a first order phase 
transition between hadronic and strange quark matter and we set $a_{2} = 
5000$ MeV$^2$.} 
\label{fig:gvsp}
\label{fig:hyb_eos}
\end{figure}

We start writing $g$ as a function of $p$ for both phases. This is 
straightforward for the SQM, solving Eq.~\eqref{eq:p(g)} for $g$, one 
obtains
 \begin{equation}
 g_{q}(p) = \left\{ \frac{9a_{2}}{2a_{4}} \left[ 1+\sqrt{1 + 
\frac{(4\pi)^2a_{4}}{3a_{2}^2}(p+B)} \right] \right\}^{1/2}.
 \end{equation}
For the chosen hadronic EoS, we construct the Gibbs free 
energy per baryon with the definition
 \begin{equation}
 g = \frac{\epsilon + p}{n_{b}}, \nonumber
 \end{equation}
from which we obtain an expression for $g_{h} \equiv g_{h}(p)$ by 
interpolating the hadronic EoS. The Gibbs condition 
\eqref{eq:gibbs-cond} can be satisfied numerically by finding the root of 
$f(p) = g_{q}(p) - g_{h}(p)$. The result is the pressure $p_T$ at which 
the two curves $g(p)$ meet (see Fig. \ref{fig:gvsp}) and corresponds to the 
constant pressure along the phase transition. The value $g_T = g(p_T)$ 
is the common Gibbs free energy per baryon. The next step is to obtain 
the energy density $\epsilon$ associated with $p_{T}$ (or $g_{T}$) for 
both EoS. In the case of SQM, from Eqs.~\eqref{eq:e} and~\eqref{eq:g}, 
we obtain
 \begin{equation}
 \epsilon_{q} = \frac{a_{4}}{36\pi^2}g_{T}^4 
-\frac{a_{2}}{12\pi^2}g_{T}^2 + B,
 \end{equation} 
which sets the energy density at the phase transition for the 
SQM phase. Similar calculations in the hadronic phase must be carried 
out to find $\epsilon_{h}$. The final result is a HyS EoS produced by 
connecting the hadronic EoS to the SQM EoS Eq.~\eqref{eq:eosq} with $p = p_{T}$ 
for $\epsilon_{h} < \epsilon < \epsilon_{q}$, as shown in Fig. 
\ref{fig:gvsp}.

\subsection{HyS EoS as piecewise polytropes}\label{subsec:HyS pwp}

For the dynamical evolution of the binary HyS with the BAM 
code~\cite{Bruegmann:2006at,Thierfelder:2011yi,Dietrich:2015iva,Bernuzzi:2016pie} 
and to calculate the initial condition using the 
SGRID code \cite{Tichy:2009yr}, we express the EoS as a piecewise polytrope (PwP):
 \begin{equation}
 p(\rho) = K_{i}\rho^{\Gamma_{i}}, \ 
d\left(\frac{\epsilon}{\rho}\right)=-p d\left(\frac{1}{\rho}\right), \ 
\rho_{i-1} < \rho < \rho_{i}\label{eq:pwp}
 \end{equation}
with  $K_{i}$ and $\Gamma_{i}$ being defined in the individual intervals 
$\rho_{0} <  ... < \rho_{i} < ... <\rho_{n}$ such that the continuity of 
$p(\rho)$
is guaranteed. 

The SLy part of our hybrid EoS uses the piecewise polytrope 
representation given in Ref.~\cite{Read:2008iy}. For the SQM phase, 
Eq.~\eqref{eq:eosq} has to be interpolated as polytropes in rest mass 
density. This is done producing a tabulated version of the SQM EoS and 
dividing this part of the EoS into fifteen pieces from which we built 
the SQM piecewise polytrope. The complete HyS piecewise 
polytrope can be found in Appendix~\ref{appA}. In 
Fig.~\ref{fig:figmvsr} we compare the results of the mass-radius curve 
for an isolated TOV star obtained using the tabulated EoS and the 
piecewise polytrope EoS. For the gravitational mass used in this 
article, we find a difference less than $0.5\%$ in the star radius.

\begin{figure}[htpb]
\centering
\includegraphics[scale=0.65]{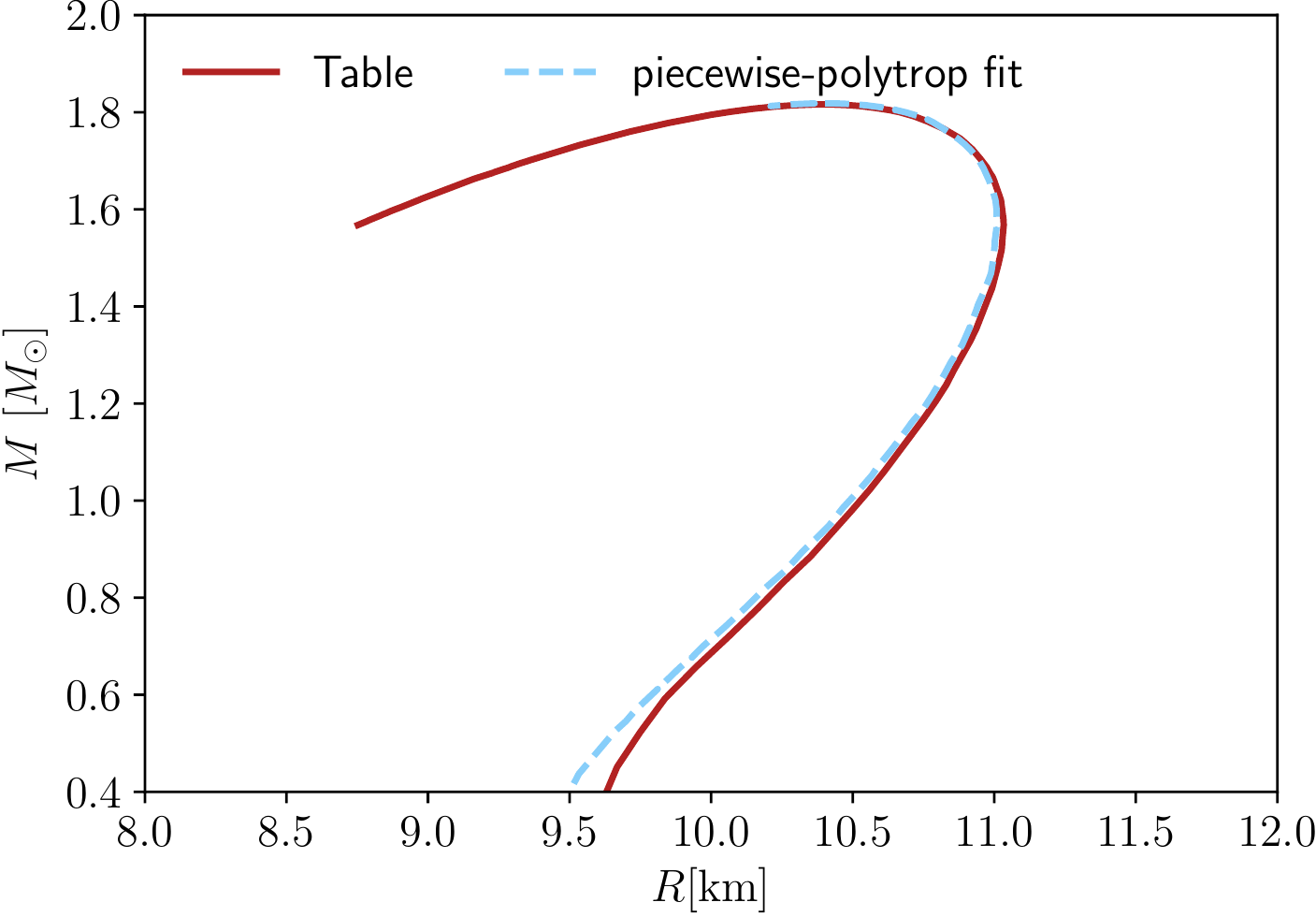}
\caption{Mass-radius curves of the tabulated and piecewise-polytrope
EoSs with $a_4 = 0.70$, $a_2 = 5000$ MeV$^2$ and $B^{1/4} = 150$ MeV. }
\label{fig:figmvsr}
\end{figure}

\section{Configurations details and numerical methods}

\subsection{Hybrid star configuration}

For our primary study of the binary HyS coalescence, 
we chose a setup which is close to the estimated binary properties of GW170817 
(details are given in Tab.~\ref{tab:config}). 
Our configuration consists of two non-spinning stars with individual masses of $M=1.365M_\odot$. 
The total mass of $M=M^A+M^B=2.73M_\odot$ is consistent with the estimated total mass of GW170817 
($2.73_{-0.01}^{+0.04}[M_\odot]$, Ref.~\cite{Abbott:2018wiz}). 
The tidal deformability of the individual stars is $\Lambda^{A,B}=378.4$ and 
the radii of the stars are $\sim 11.0 \rm km$, which also agrees with current multi-messenger 
constraints arising from the analysis of the GW, 
the kilonova, and the short GRB~\cite{Coughlin:2018fis} of $R \in [10.9,13.6]\rm km$.

\begin{table}[htpb]
\caption{Properties of the employed configuration. 
The columns refer to the individual masses in isolation, 
the baryonic mass of the individual stars, the stars' compactnesses, 
the radii of the stars, the tidal
deformability of the individual stars ${\Lambda}^{A,B}$,  
the ADM-mass and angular momentum of the system, and the estimated 
residual eccentricity $e$.}
\label{tab:config}
\centering
\begin{tabular}{ccccccccc}
\hline 
$M^{A,B} [M_\odot]$ & $M_b^{A,B} [M_\odot] $ & ${C}^{A,B}$ & $R^{A,B}\ [\rm km]$ & 
${\Lambda}^{A,B}$ & $M_{\rm ADM} [M_\odot]$& $J_{\rm ADM} [M_\odot^2]$ & $e$ \\
\hline
1.365 & 1.527 & 0.18 & 11.0 & 378  & 2.705 & 8.324 & $5\times 10^{-3}$\\
\hline 
\end{tabular}
\end{table}

\subsection{Numerical methods}

We compute the initial configuration solving the 
conformal thin-sandwich equations~\cite{Wilson:1995uh,Wilson:1996ty,York:1998hy}
with the SGRID code~\cite{Tichy:2009yr,Tichy:2012rp,Dietrich:2015pxa}. 
SGRID uses pseudospectral methods to accurately compute 
spatial derivatives. 
We performed an iterative procedure to reduce the orbital 
eccentricity varying the system's initial radial velocity and 
eccentricity parameter, as discussed 
in~\cite{Moldenhauer:2014yaa,Kyutoku:2014yba,Dietrich:2015pxa}, 
until eccentricity is $\sim$ $10^{-3}$.

Dynamical evolutions are performed with the BAM 
code~\cite{Bruegmann:2006at,Thierfelder:2011yi,
Dietrich:2015iva,Bernuzzi:2016pie}, which  
uses the method of lines employing 
finite differences for the spatial discretization of the metric variables 
and a high-resolution shock-capturing (HRSC) scheme for the computation of the 
matter fluxes of the general relativistic hydrodynamics equations. 
The spacetime is evolved with the Z4c evolution system 
\cite{Bernuzzi:2009ex,Weyhausen:2011cg,Hilditch:2012fp}. 
The BAM code employs an adaptive mesh refinement strategy with 
a hierarchy of nested Cartesian 
grids. The grid spacing of each level is half the grid spacing of 
its coarser parent level. 
A number of inner levels can be moved dynamically to cover 
the NS's during the evolution. 
In this article, we use $7$ refinement levels labeled 
$l = 0,...,6$~ from the coarsest to the finest. 
The detailed grid specifications for the employed resolutions 
are given in Tab.~\ref{tab:NRgrid}. 
The GRHD numerical fluxes are computed with a local Lax-Friedrich's (LLF) scheme 
and the fluid's characteristic variables are reconstructed using a fifth order 
weighted essentially non-oscillatory (WENOZ) algorithm \cite{Borges:2008a,Bernuzzi:2016pie}.
The same GRHD settings have been employed in~\cite{Bernuzzi:2016pie,Dietrich:2017aum,
Dietrich:2018upm} leading to a 2nd order convergence of the GW phase independent 
of the configuration details. 

\begin{table}[h]
\caption{Grid configurations employed in the simulations for the different resolutions. 
The columns refer to the simulation's name,  
the number of moving boxes $L^{\rm mv}$, the number of 
grid points per direction in the static levels $n$, the number of grid points
per direction in the moving levels $n^{\rm mv}$, 
the grid spacing in the innermost level $h_{6}$ (corresponding to $l = 6$), and 
the grid spacing of the coarsest level $h_0$.}
\label{tab:NRgrid}
\centering
\begin{tabular}{c|ccccc}
\hline 
Name & $L^{\rm mv}$ & $n$ & $n^{\rm mv}$ & $h_{6}$ & $h_0$ \\
\hline
R1 & $4$ & $192$ & $64$ &  $0.240$ & $15.36$ \\
R2 & $4$ & $288$ & $96$ &  $0.160$ & $10.24$ \\
R3 & $4$ & $384$ & $128$ & $0.120$ & $7.68$  \\
R4 & $4$ & $480$ & $160$ & $0.096$ & $6.14$  \\
\hline 
\end{tabular}
\end{table}

\section{Quantitative Merger Dynamics}

\begin{figure}[htpb]
\centering
\includegraphics[width=0.495\textwidth]{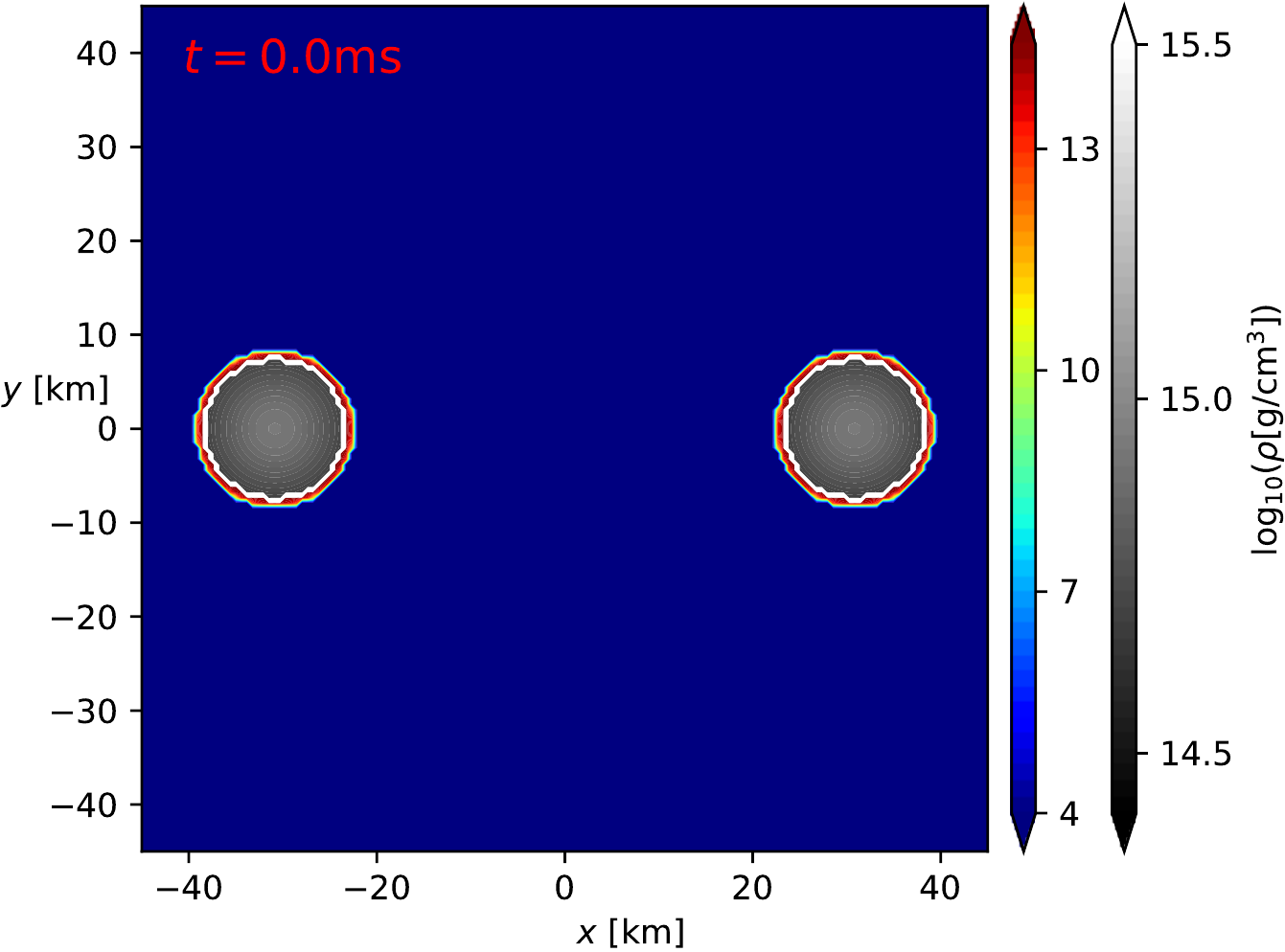}\hfill
\includegraphics[width=0.495\textwidth]{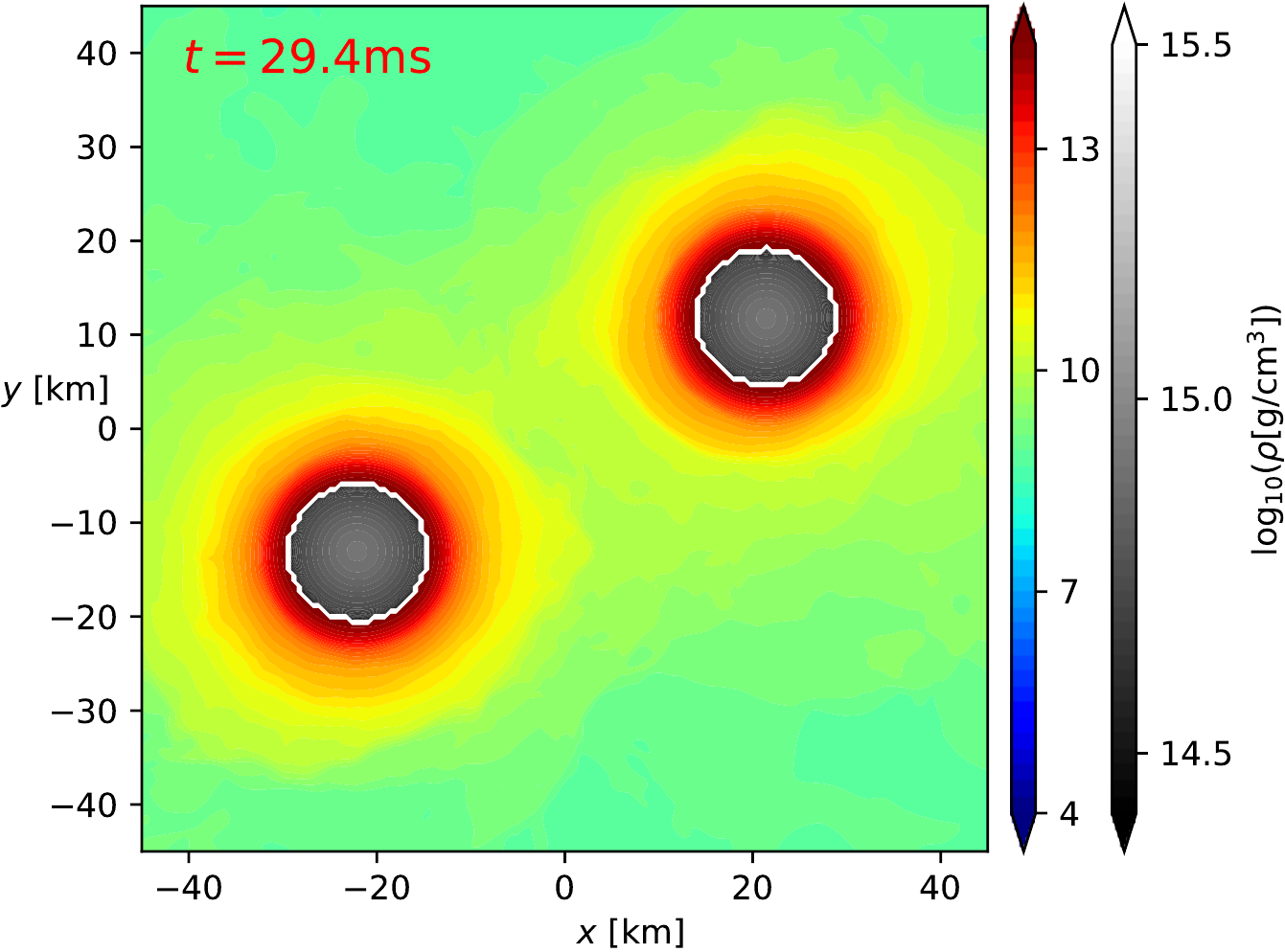}\\
\includegraphics[width=0.495\textwidth]{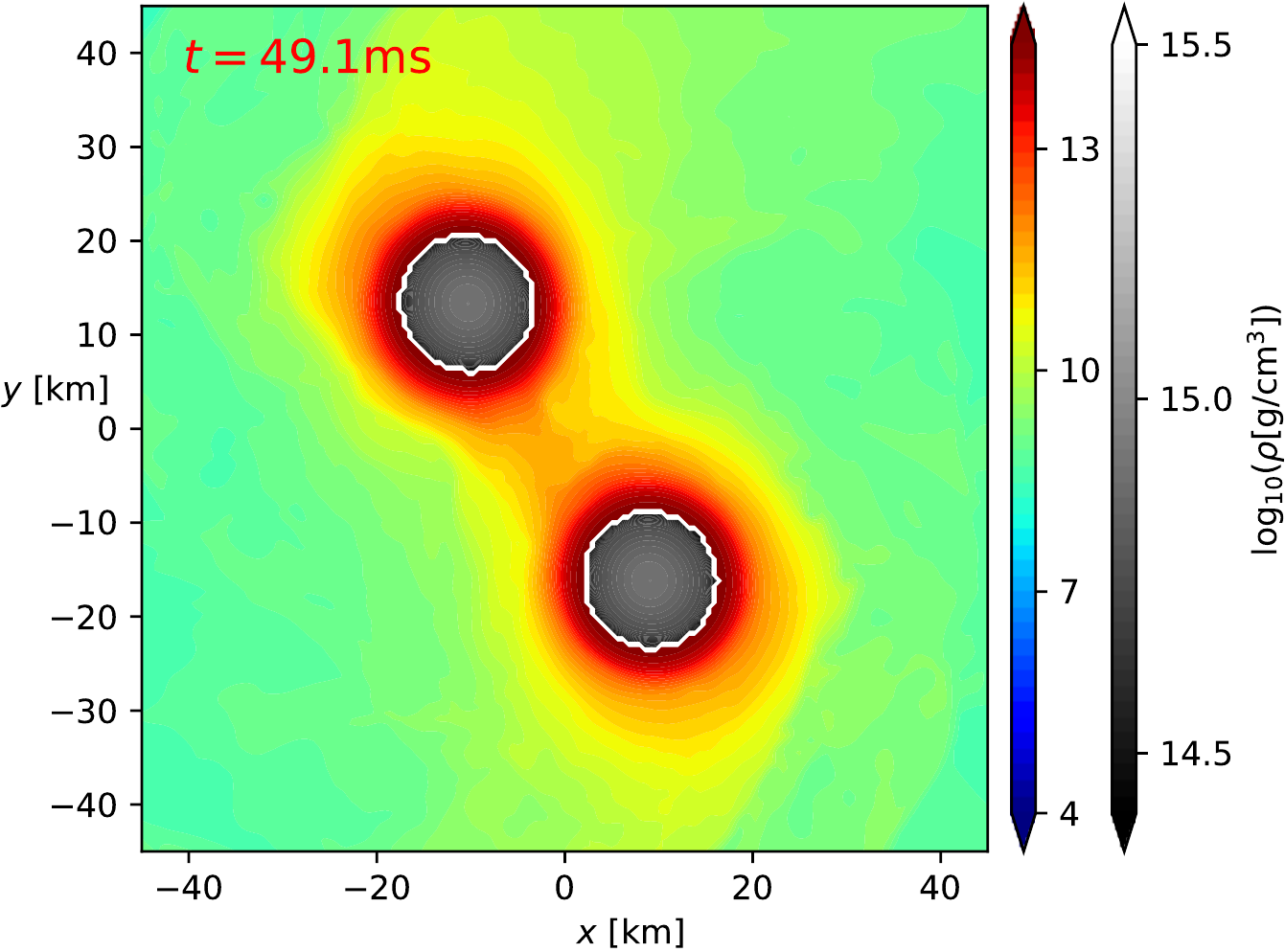}\hfill
\includegraphics[width=0.495\textwidth]{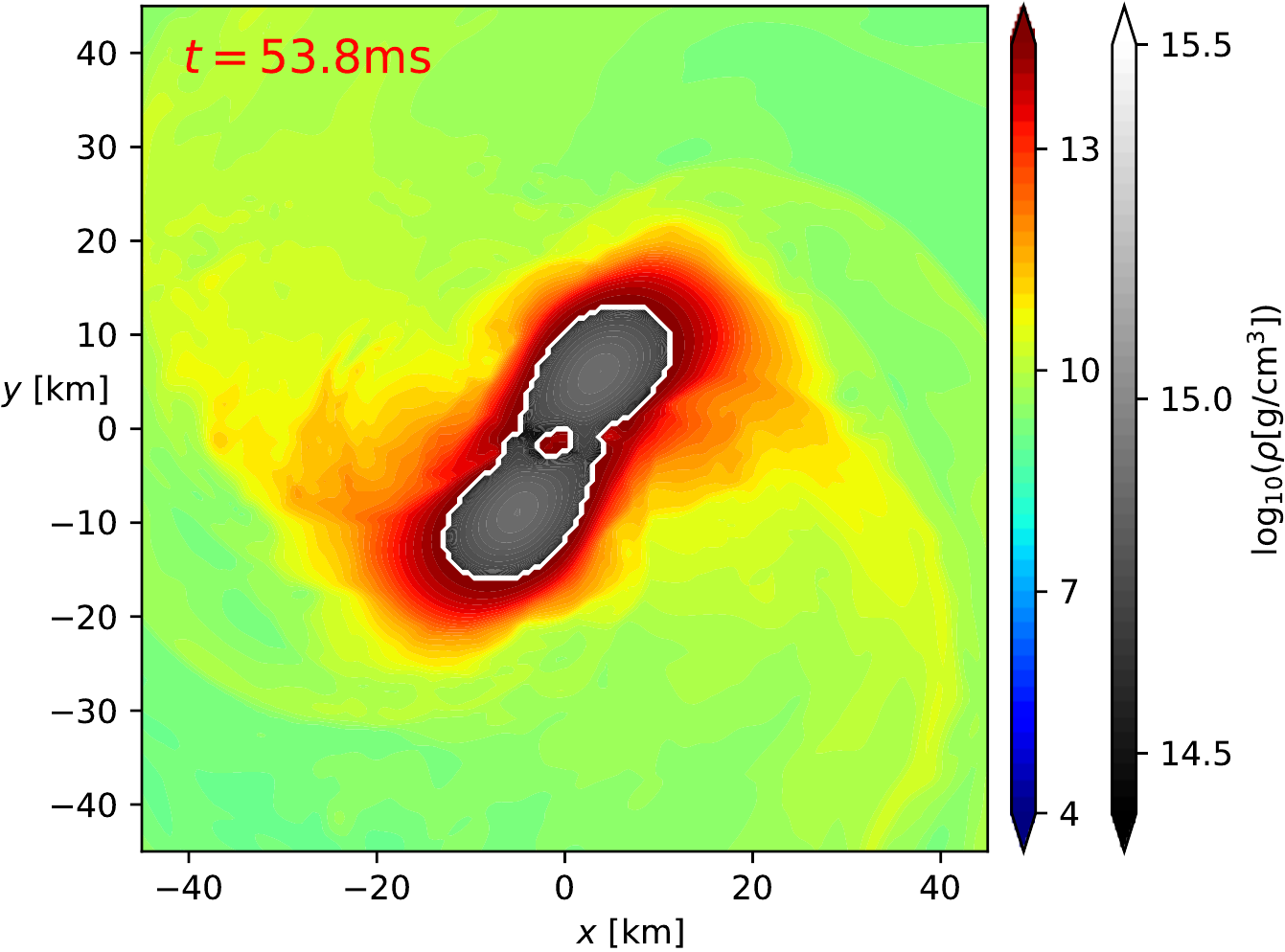}\\
\includegraphics[width=0.495\textwidth]{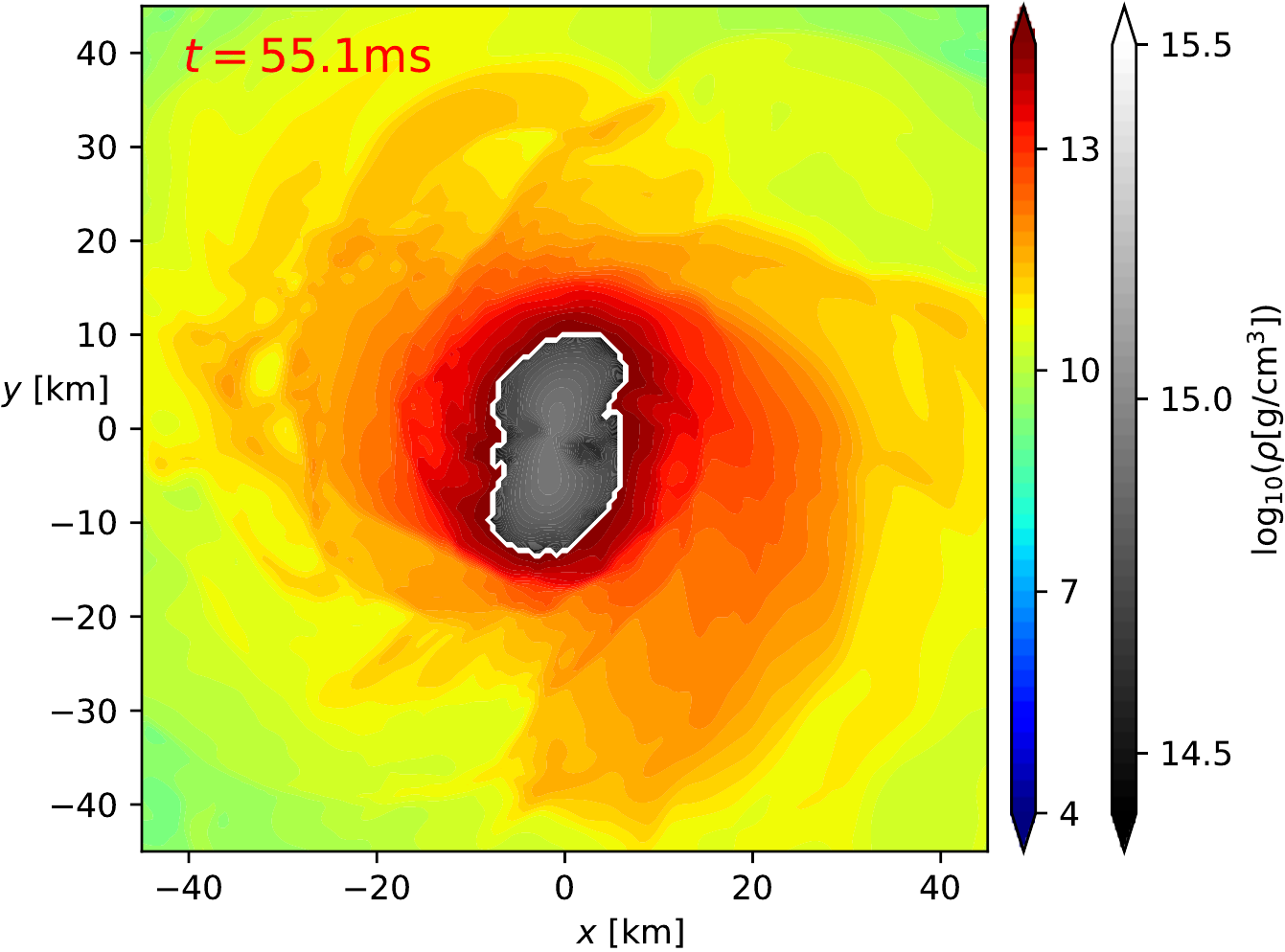}\hfill
\includegraphics[width=0.495\textwidth]{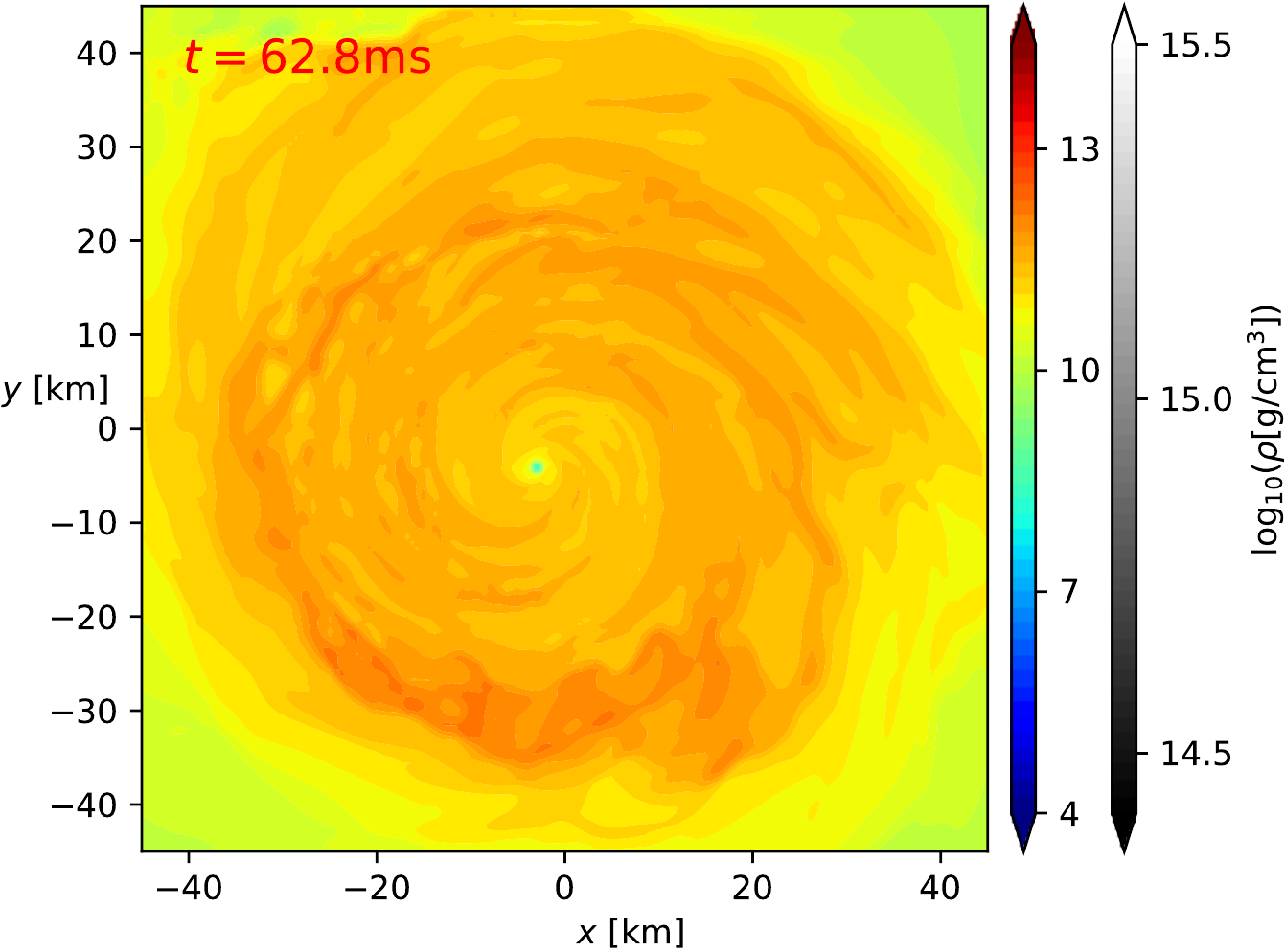}
\caption{Density evolution within the orbital plain for different time snapshots for the 
R4 setup. We show the HM on a color scale ranging from 
blue to red and the SQM on a gray scale.}
\label{fig:density}
\end{figure}

We start discussing our binary HyS mergers with a qualitative description of the 
coalescence of the two stars as depicted in Fig.~\ref{fig:density}. 
The first panel shows the initial setup for our evolution as computed from SGRID. 
The two stars have an initial coordinate separation of $\sim 61.5\rm km$, 
which corresponds to a proper distance of $\sim 76.4\rm km$. 
As visible in the figure, we use a gray scale for the SQM, 
while HM is shown on a blue to red scale. 
The second panel shows the system after approximately 6 orbits, i.e., 12 GW cycles. 
At this time one sees that the star's surface gets `smeared out' 
and is not perfectly preserved. Thus, lower density material surrounds 
the two stars. 
Similar observations are typical for full NR simulations and have been 
discussed in e.g.~\cite{Faber:2012rw} and references therein. 
The main reason for this, is the sharp transition of the density and 
the usage of an artificial atmosphere for the GRHD 
simulation, see e.g.~\cite{Bernuzzi:2016pie}. 
A similar observation is true about three to four orbit before the moment of merger (third panel). 
The fourth panel shows the system at the merger, i.e., when the GW amplitude peaks. 
One sees that the two SQM cores of the individual stars come into contact at this time. 
After the moment of merger, the central density inside the SQM core increases.
The formed remnant (fifth panel of Fig.~\ref{fig:density}) survives for 
about $\sim 5$ milliseconds and undergoes two oscillations (three oscillations for lower resolutions) 
before it collapses to a black hole (BH).
The final system (sixth panel) consists of a BH surrounded by 
an accretion disk composed of HM. 

\section{Gravitational Wave Signal}

\begin{figure}[htpb]
\centering
\includegraphics[width=0.95\textwidth]{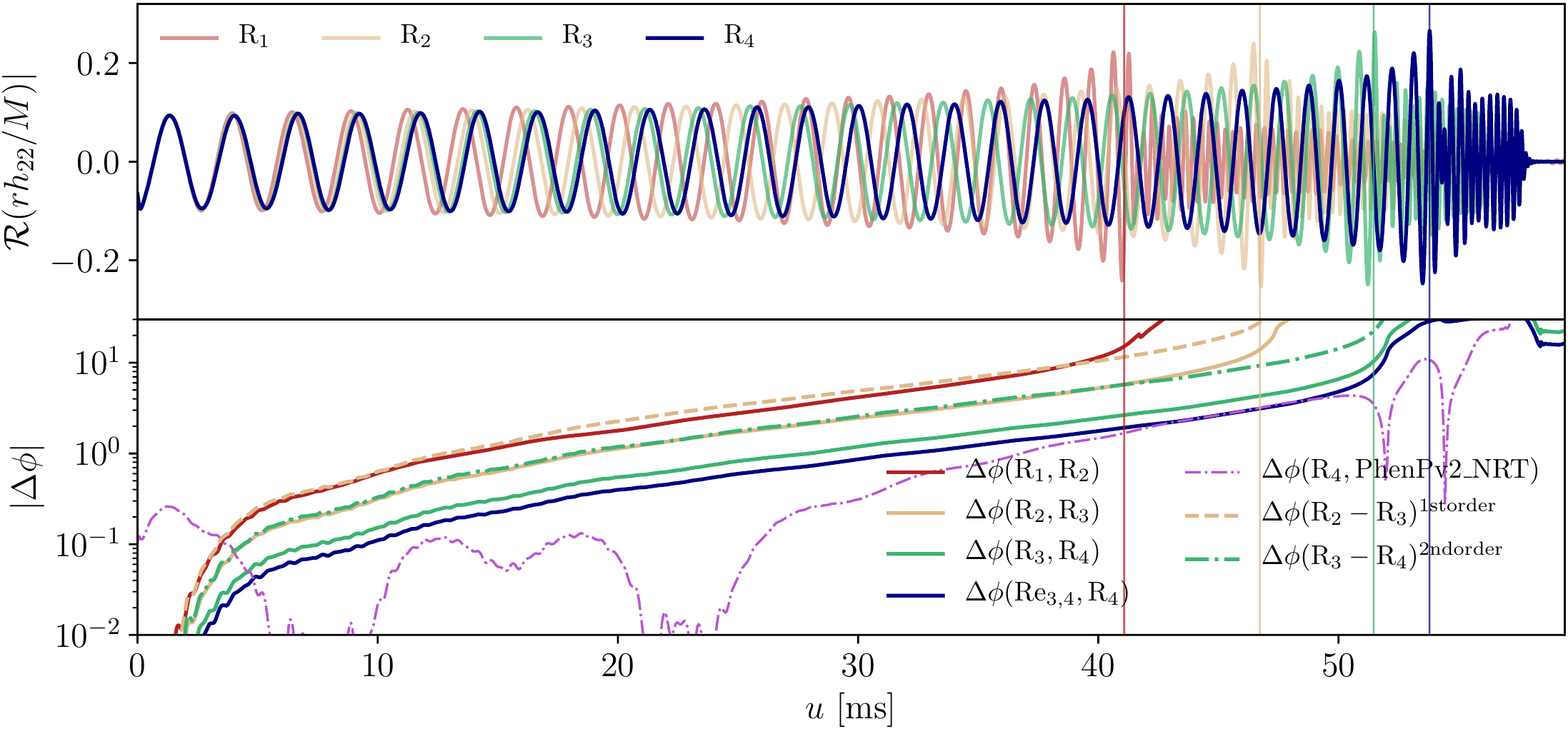}
\caption{The top panel shows the (2,2)-mode of the GW signal for the different resolutions. 
The bottom panel shows the phase difference between the individual resolutions, where dashed lines
show the phase difference rescaled to an assumed first or second order convergence. 
In addition, we show the phase difference with respect to a Richardson extrapolated waveform using resolutions 
$\rm R3$ and $\rm R4$, denoted by $\rm Re_{3,4}$. 
We also include the phase difference of $\rm Re_{3,4}$ 
with respect to the waveform approximant 
$\texttt{IMRPhenomPv2\_NRTidal}$~\cite{Dietrich:2018uni} (purple dash-dotted line), 
denoted as \texttt{PhenPv2\_NRT}.
Vertical lines mark the moments of merger for the individual resolutions.}
\label{fig:convergence_phi}
\end{figure}

\subsection{The Inspiral}

We present the full gravitational waveform obtained from our simulations 
in Fig.~\ref{fig:convergence_phi} (top panel). Quantities related
to the GW signal are plotted against the retarded coordinate time $u$,
defined as
\begin{equation}\label{eq:u}
u = t - r_{\rm ext} -2M\ln(r_{\rm ext}/2M - 1),
\end{equation}
where we choose $r_{\rm ext}=1000M_\odot$.
The inspiral is characterized 
by an increasing frequency and amplitude of the GW, the so-called chirp-signal. 
We find that the merger time (marked by a vertical line) shifts by several milliseconds 
between the different resolutions. It is important to point out that for similar resolutions, 
cf.~\cite{Dietrich:2016hky,Dietrich:2016lyp}, the dephasing and consequently 
the shift in the merger time between similar resolutions have been significantly 
smaller for EoSs without phase transition. 
The bottom panel of Fig.~\ref{fig:convergence_phi} shows the phase differences 
between individual resolutions. 
One finds that for the three lowest resolutions, the phase is only converging with first order 
with respect to the grid spacing. Rescaling the phase differences assuming first order convergence
leads to very good agreement with the phase difference of lower resolutions (dashed yellow line). 
This is of particular importance since it has been pointed out that the combination of 
the high-order flux reconstruction for the characteristic variables with the 5th 
order WENOZ scheme leads to a clear second-order convergence 
of the GW phase~\cite{Bernuzzi:2016pie,Dietrich:2017aum,Dietrich:2018upm} 
for a variety of hadronic EoSs employing the same resolution and code settings.
Increasing the resolution, we find that second-order 
convergence is recovered (dash-dotted green line). 
Ref.~\cite{Bernuzzi:2016pie} pointed out that the NS's surface causes this 
second-order convergence even though the numerical scheme is 5th order convergent for 
smooth hydrodynamical problems.
Here, we find that in the case of a first order phase transition inside the NSs, the 
convergence order can even drop further to first order, but that with higher resolutions 
the convergence order increases.
The reduced convergence order explains the large dephasing between 
the individual resolutions and displays the additional challenges in 
NR simulations of NS binaries employing a hybrid EoS.

Further work employing different hybrid EoSs and numerical schemes (varying the 
numerical flux scheme and limiters) is needed to understand if the observed feature 
is universal or might be caused by a particular choice of our parameters. \\

Finally, assuming second order convergence, we perform a Richardson 
extrapolation to obtain a better prediction for the GW phase, denoted by 
$\rm Re_{3,4}$. We compare $\rm Re_{3,4}$ with the GW approximant 
\texttt{IMRPhenomPv2\_NRTidal}~\cite{Dietrich:2018uni,Dietrich:2017aum,Hannam:2013oca}, 
which has been employed by the Ligo and Virgo Collaborations for the 
analysis of GW170817 
in~\cite{Abbott:2018wiz,Abbott:2018exr,LIGOScientific:2018mvr,Abbott:2018lct}. 
This approximant (\texttt{IMRPhenomPv2\_NRTidal}) is a 
phenomenological waveform model which has been developed for the 
description of BNS systems. During its construction only NR simulations 
with hadronic EOSs were used~\cite{Dietrich:2017aum}. Therefore, we 
use this model to compare our results of hybrid star binaries with 
predictions for purely hadronic EOSs. The exact waveform 
\texttt{IMRPhenomPv2\_NRTidal}, which we compare to, uses the same 
masses and tidal deformabilities as our NR simulation. Finally, the 
\texttt{IMRPhenomPv2\_NRTidal} and our hybrid star waveform are aligned 
within a time window $u\in[5,10]\rm ms$ by minimizing the phase 
difference within this interval to allow a reasonable comparison and to
compute the phase difference. The phase difference between resolution 
$\rm Re_{3,4}$ and \texttt{IMRPhenomPv2\_NRTidal} is shown as a 
dash-dotted purple line in Fig.~\ref{fig:convergence_phi} (bottom 
panel). We find that until the moment of merger, the phase difference 
$\phi ({\rm Re_{3,4}}) - \phi({\rm IMRPhenomPv2\_NRTidal})$ is below the 
estimated error $\Delta \phi (\rm Re_{3,4},R_4)$~\footnote{Note that the 
larger dephasing at the beginning of the simulation can be explained by 
junk radiation.}, which is the first validation of a waveform model 
against a NR dataset including a strong phase transition inside the 
star. That the difference between the NR data and the waveform 
approximant is below the uncertainty of the data strengthens the 
reliability of models used for GW data analysis even if 
these have been derived for purely hadronic EOSs. However, due to the 
reduced convergence order and the large phase error of our simulations, 
further evolutions with higher resolutions are required for a more 
quantitative and stronger test.

\subsection{The Postmerger}

After the merger of the two stars, the rotating merger remnant emits GWs due to its 
asymmetric shape (see the fifth panel of Fig.~\ref{fig:density}). 
The typical emission during the postmerger happens on frequencies ranging from $2-4\rm kHz$. 
Which frequencies are excited and the strength of each individual frequency 
depends on the properties of the merger remnant and, thus, on the 
mass, spin, and EoS of the binary. 
We refer to existing studies available in the 
literature, e.g.,~\cite{Takami:2014zpa,Takami:2014tva,Rezzolla:2016nxn,
Bauswein:2011tp,Stergioulas:2011gd,Bauswein:2012ya,Bauswein:2014qla,
Clark:2015zxa,Bauswein:2015yca,Bernuzzi:2015rla,Dietrich:2016lyp,
Dietrich:2016hky,Dietrich:2015iva}, for further details and discussions about 
the mechanisms causing GW emission during the postmerger evolution.
However, we want to understand if the presence of a SQM core changes 
the standard picture of the characteristic postmerger spectrum estimated for 
hadronic matter, see also~\cite{Bauswein:2018bma,Most:2018eaw}.

\begin{figure}[htpb]
\centering
\includegraphics[width=0.95\textwidth]{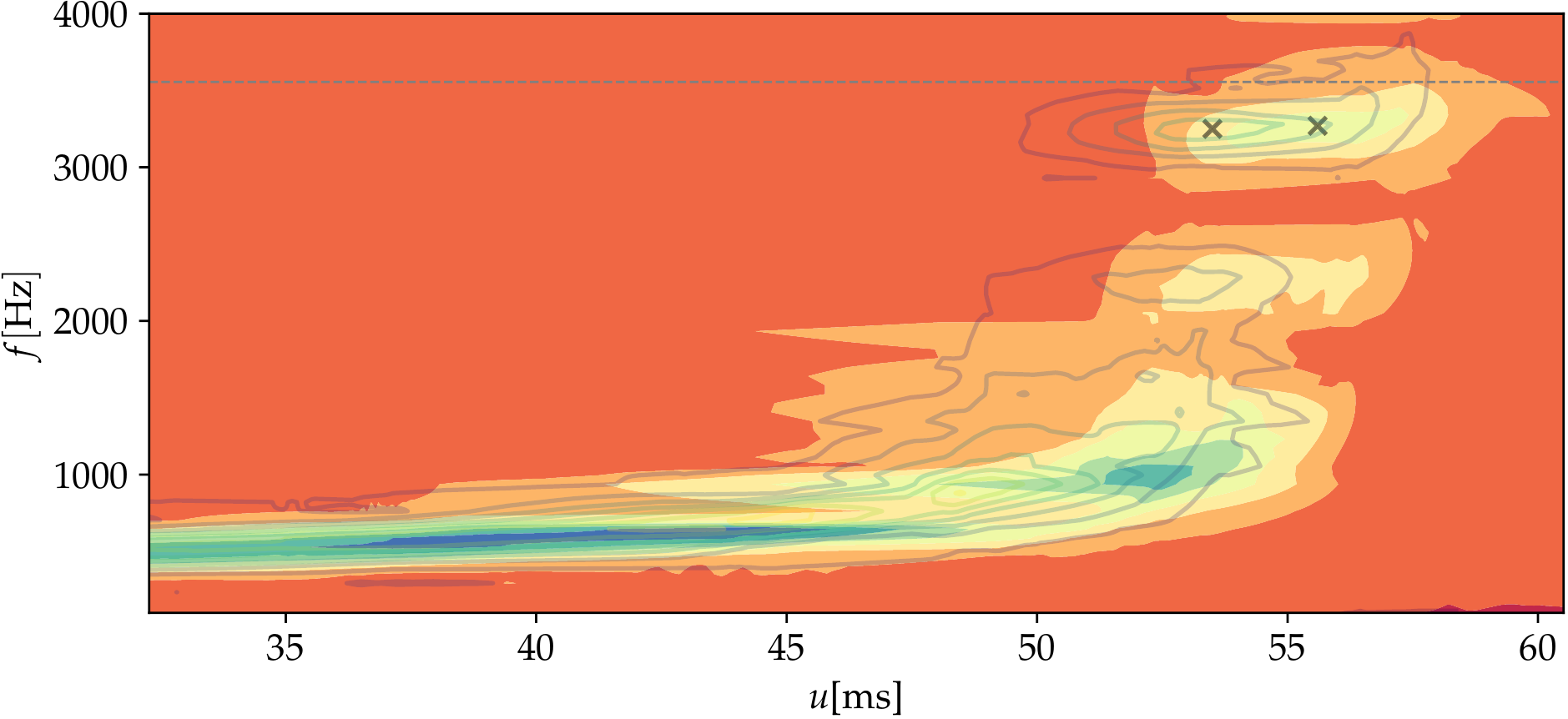}
\caption{Spectrogram for resolution R4 (filled contours) and resolution R3 (solid lines). 
We mark the dominant postmerger frequency for the two resolutions as black crosses.
In addition we include as a horizontal dashed line the postmerger frequency estimate 
derived from Ref.~\cite{Bernuzzi:2015rla}.}
\label{fig:spectrogram}
\end{figure}

For a better understanding of the frequency evolution, we present the spectrogram 
of the dominant (2,2)-mode of the GW signal in Fig.~\ref{fig:spectrogram}; 
we refer to Ref.~\cite{Chaurasia:2018zhg} for a detailed discussion about the 
computation of the spectrogram. 
For our setup, the merger remnant collapses within $\sim 5\ \rm ms$ to a BH, 
due to this very short time the postmerger GW emission is not very strong. 
Nevertheless, one can extract the main emission frequency $f_2$ and obtains an emission of 
\begin{equation}
 f_2 = ( 3.27 \pm 0.1) \ \rm kHz, 
\end{equation}
where the uncertainty is determined as the difference 
between the two highest available resolutions and also includes the uncertainty 
caused by the finite width of the peak in the frequency domain spectrum. 
We compare this $f_2$ frequency with the quasi-universal 
relation expressed in Ref.~\cite{Bernuzzi:2015rla} derived from pure hadronic EoSs. 
For a combined tidal deformability of $\tilde{\Lambda}=378$~\footnote{The
combined tidal deformability $\tilde{\Lambda}$ equals the individual 
tidal deformabilities $\Lambda^{A,B}$ presented in Tab.~\ref{tab:config}
for equal-mass binary stars.} 
Ref.~\cite{Bernuzzi:2015rla} predicts 
$f_2=3.55\rm kHz$. 
We visualize this estimate as a horizontal dotted 
line in Fig.~\ref{fig:spectrogram}. 

Overall, the difference of $0.3\rm kHz$ between our measurement and the quasi-universal 
relation is consistent with the uncertainty of the phenomenological fit~\cite{Bernuzzi:2015rla}
combined with the uncertainty of our NR data. 
Thus, no clear imprint of the SQM core of the merger remnant 
on the postmerger evolution is present. This is opposite to the work 
of~\cite{Bauswein:2018bma} in which the merger remnant
undergoes a first order phase transition which changes significantly the postmerger 
spectrum leading to a main emission frequency about $450\\rm Hz$
higher than the one predicted for a pure hadronic EoS. 
However, since the SQM imprint on the postmerger dynamics will depend on the 
exact EoS, our results are not in contradiction or tension to~\cite{Bauswein:2018bma}. 
Similar to our results Ref.~\cite{Most:2018eaw} finds 
a shorter merger remnant lifetime in the presence of 
a phase transition,  
but a similar postmerger frequency evolution as for pure hadronic matter. 

\section{Ejecta and electromagnetic counterparts}

In addition to the emission of GWs, NS mergers are associated with a variety 
of EM signals such as short gamma-ray bursts (GRB), a synchrotron afterglow, 
and a thermal kilonova. 

\begin{figure}[htpb]
\centering
\includegraphics[width=0.98\textwidth]{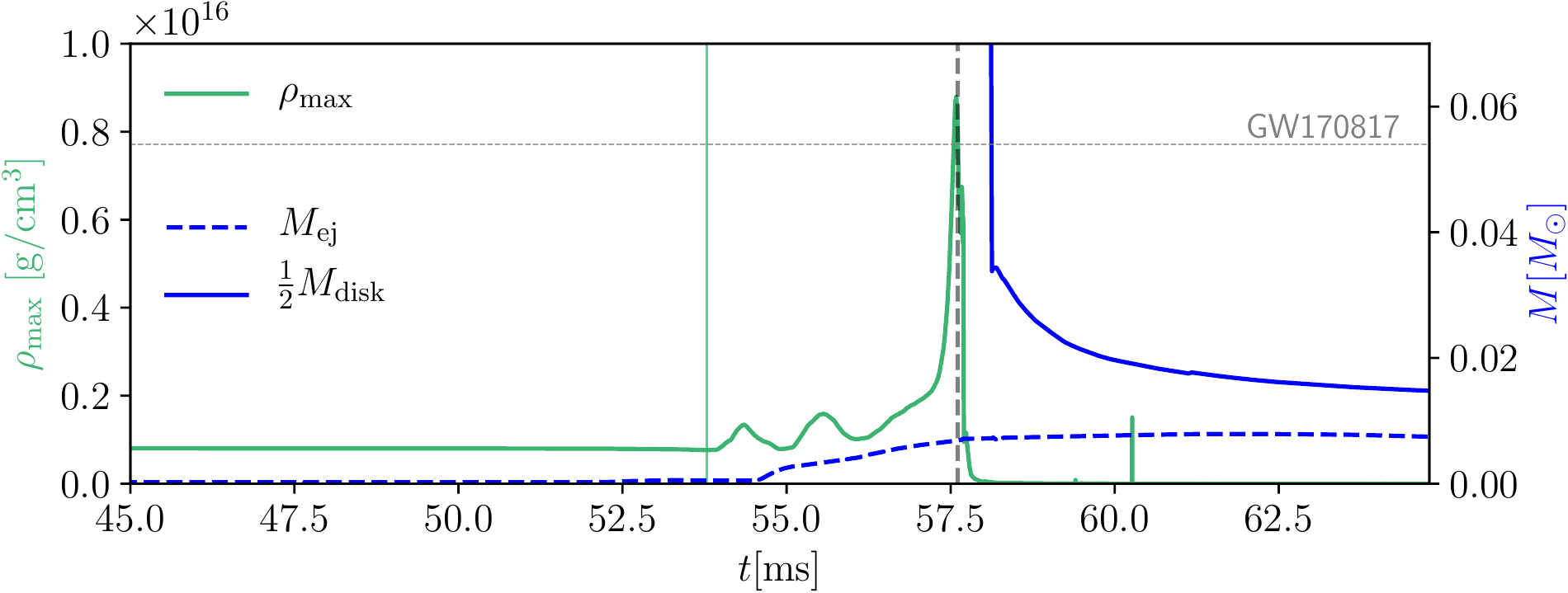}
\caption{Time evolution of the maximum density (green) and the amount of 
ejected material (dashed blue) and the disk mass (solid blue) for 
resolution R3. The ejected material is estimated 
according to Eqs.~(14) and (15) of \cite{Dietrich:2015iva}, i.e., all 
material is marked as ejecta as long as it is gravitationally unbound and 
has an outward pointing velocity.}
\label{fig:rho_max}
\end{figure}

The thermal kilonova originates from neutron-rich outflows and 
the radioactive decay of $r$-process elements in the ejected matter, 
see e.g.~\cite{Metzger:2016pju} and references therein. 
Kilonovae produce an almost isotropic EM emission and are 
visible in the ultraviolet/optical/infrared band. 
Current analyses of the kilonova AT2017gfo associated with GW170817
estimate the ejecta mass to be $\sim 5\times 10^{-2}M_\odot$. 
In our work, we will assume $\sim 5.4\times 10^{-2}M_\odot$ 
as found in~\cite{Coughlin:2018fis}. 

We present the amount of ejected matter present over the course of our 
simulation in Fig.~\ref{fig:rho_max} as a blue dashed line. 
We find that after the merger, the maximum density of the NS remnant increases 
and 3 oscillations are found. 
During this time most of the ejection happens and at the end of our simulation 
the ejecta mass is about $\sim 0.8\times 10^{-2}M_\odot$. 
This estimate agrees to about $30\%$ with resolution R3, 
which shows the overall consistency of our dataset. 
For comparison, the ejecta mass obtained in equal-mass BNS 
simulations employing SLy EoS with similar grid configurations and masses 
(Tab. III of Ref.~\cite{Dietrich:2016hky}, 
with $M^{A,B} =  1.37 M_{\odot}$ and Tab. V of Ref.~\cite{Dietrich:2015iva}, 
with $M^{A,B} = 1.35 M_{\odot}$) are, respectively, 
$1.6\times10^{-2} M_{\odot}$ and $1.22\times10^{-2} M_{\odot}$. 
Consequently, our binary HyS configuration ejects only half of the 
amount of dynamical ejecta mass when compared to simulations 
employing the purely hadronic EOS 
SLy~\cite{Dietrich:2016hky,
Dietrich:2015iva}~\footnote{We note that the phenomenological 
fit for hydonic EOSs for the dynamical ejecta mass proposed 
in~\cite{Coughlin:2018fis} (which is an updated version of the work 
presented in~\cite{Dietrich:2016fpt})
predicts a dynamical ejecta mass of $0.28\times 10^{-2}M_\odot$ , 
which lies even below our results. The discrepancy between the fit 
and the SLy simulations is caused by the large shock driven 
ejecta for the SLy EOS, which seems not to be fully 
captured in~\cite{Coughlin:2018fis}. In addition, this comparison shows 
that further work with a larger set of HyS EOSs is required to obtain further 
confidence in the ejecta properties of NR simulations.}.

Due to the short evolution time after the merger, we are unable to estimate the total amount 
of ejecta due to the missing disk wind driven outflows which happen on longer timescales, 
see e.g.~\cite{Kohri:2005tq,
Surman:2005kf,Metzger:2008av,Dessart:2008zd,Fernandez:2013tya,
Perego:2014fma,Siegel:2014ita,Just:2014fka,
Rezzolla:2014nva,Ciolfi:2014yla,Siegel:2017nub}. However, we find that the disk 
mass drops to about $\sim 3.4\times 10^{-2}M_\odot$ 
towards the end of our simulation. 
Compared to the SLy simulations in Refs.~\cite{Dietrich:2016hky,Dietrich:2015iva} this 
is only about $\sim 15$ to $40\%$ of the SLy simulations.
This means that if even half of the disk would be ejected (see Fig.~\ref{fig:rho_max}), 
the kilonova associated with our studied HyS configuration would not be as bright as 
it was observed for GW170817.\\

Overall, the comparison of the ejecta and disk mass of 
our binary HyS and an SLy BNS suggests that HyS systems retain more mass 
within the stars along the time evolution and create dimmer EM 
counterparts. Further studies are necessary in order to understand 
whether this behavior is universal or restricted to our SQM EoS 
parameters choice.

\begin{figure}[htpb]
\centering
\includegraphics[width=0.98\textwidth]{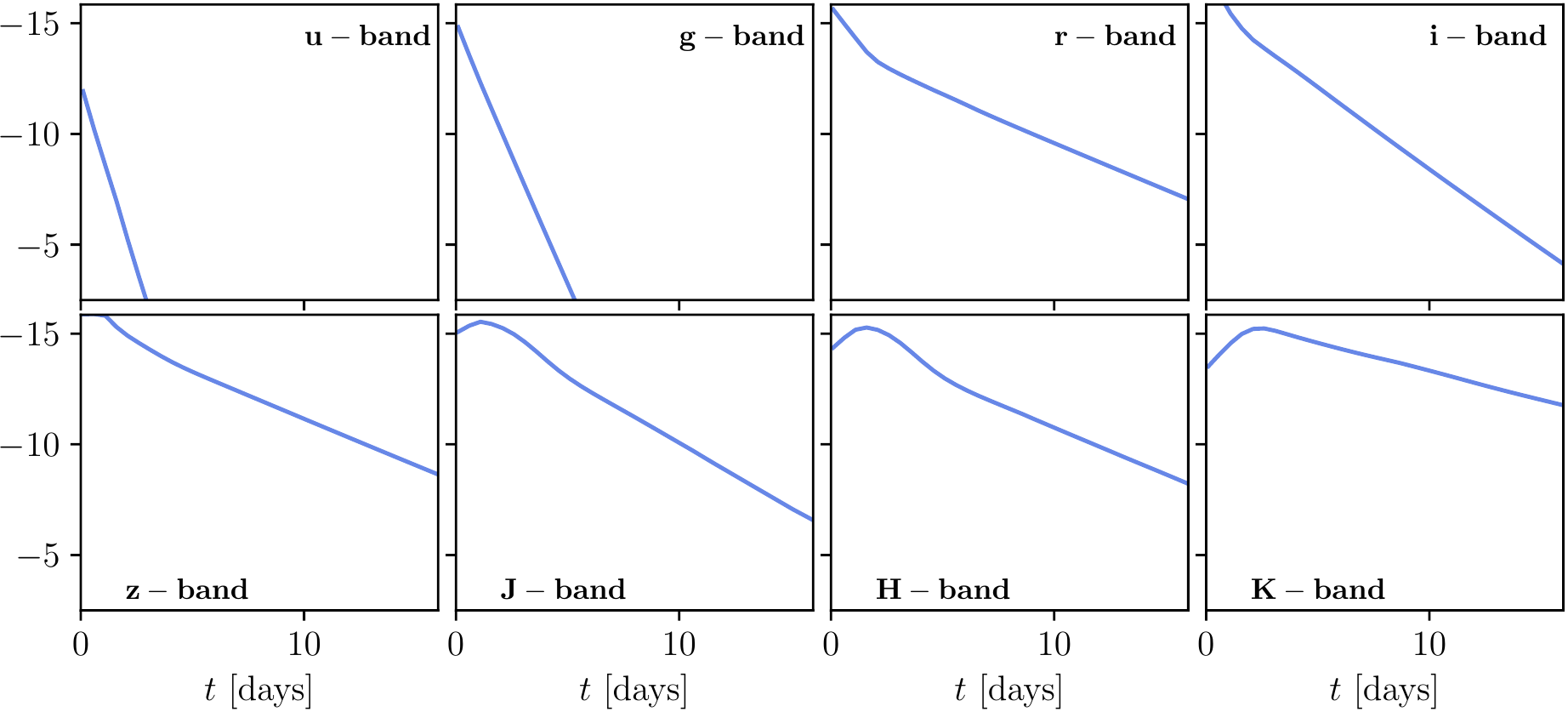}
\caption{Estimated absolute magnitude of the kilonova in different frequency 
         band using the methods outlined 
in~\cite{Kasen:2017sxr,Coughlin:2018miv}. }
\label{fig:lightcurves}
\end{figure}

To estimate the absolute magnitude of the kilonova connected to the 
HyS merger, we use the kilonova model of~\cite{Kasen:2017sxr,Coughlin:2018miv}. 
The model of~\cite{Kasen:2017sxr} employs a multi-dimensional 
Monte Carlo code to solve the multi-wavelength radiation 
transport equation for a relativistically expanding medium. 
To obtain the final predicted lightcurves, 
we interpolate between the existing Monte Carlo simulations 
using the Gaussian Process Regression techniques discussed 
in~\cite{Coughlin:2018miv}. 
For our estimates, we use the dynamical ejecta component 
$M_{\rm ej,dyn} = 8 \times 10^{-3}M_\odot$ which 
leaves the system with an average velocity of 
$v_{\rm ej,dyn} =0.2$. 
Furthermore, we assume a lanthanide fraction $X_{\rm ej, dyn}=10^{-1}$. 
For the disk wind ejecta, we assume a velocity of 
$v_{\rm ej,wind} =0.1 c$, a lanthanide fraction of $10^{-4}$, 
and an ejecta mass of $M_{\rm ej, wind} = 0.017 M_\odot$. 
The largest absolute magnitudes are obtained in the near-infrared, 
e.g.~i- and K-band, while in the optical and ultraviolet 
the signal is fainter, see Fig.~\ref{fig:lightcurves}. 

In contrast to the kilonova, short GRBs are created by highly relativistic outflows, 
powered by the merger remnant accretion disk, e.g.,~\cite{Paczynski:1986px,Eichler:1989ve}. 
Based on the measured accretion disk mass of $\sim 3\times 10^{-2}M_\odot$ at the end of 
our simulation, we can obtain an order of magnitude estimate for the energy which could be 
released by a potential short GRB. Assuming that the GRB energy is proportional to the disk mass
and that the proportionality constant is of about $10^{-2}$ as estimated in~\cite{Coughlin:2018fis}, 
we find a GRB energy of $3 \sim 10^{-4}M_\odot$, i.e., $6\times 
10^{50}$~erg, which 
is consistent with current observational constraints for GRB170817A, 
e.g.,~\cite{vanEerten:2018vgj}. 

\section{Conclusion}
In this work, we presented and discussed the general procedure to construct HyS EoSs
combining the tdBag model for SQM and a HM EoS for low-density material. 
Our procedure can be applied to study a variety of parameters and HM EoS choice and 
enables future studies of NSs binary containing SQM cores.
The GW signal obtained from our NR simulations of a non-spinning, 
equal mass binary HyS merger exhibits a larger dephasing caused by numerical discretization 
than similar simulations for purely hadronic BNS configurations. 
We point out that the presence of a strong phase transition inside the stars 
can lead to a drop of the convergence to 1st order for insufficient resolutions
even though higher order methods are applied. 
This is evidence that HyS mergers add additional complexity 
to the NR simulations. Further simulations 
must be done in order to understand if these features are universal or
caused by the particular choice of EoS in connection with the employed numerical methods.
Furthermore, we show that the postmerger GW main emission frequency for our constructed EoS is 
consistent with quasi-universal relations derived from hadronic EoSs within the uncertainties 
of the relation.
Performing an analysis of the ejecta properties and EM signatures, the kilonova and the short GRB, 
we are able to draw the full multi-messenger picture for our HyS merger simulation. 
Our work is the first step towards a better coverage of the BNS parameter space and enables future 
investigations of a wider range of EoS with full NR simulations.

\bibliography{paper20190808.bib}

\begin{thebibliography}{-------}
\providecommand{\natexlab}[1]{#1}

\bibitem[Abbott \em{et~al.}(2017{\natexlab{a}})Abbott
  et~al.]{TheLIGOScientific:2017qsa}
Abbott, B.P.; others.
\newblock {GW170817: Observation of Gravitational Waves from a Binary Neutron
  Star Inspiral}.
\newblock {\em Phys. Rev. Lett.} {\bf 2017}, {\em 119},~161101,
  \href{http://xxx.lanl.gov/abs/1710.05832}{{\normalfont
  [arXiv:gr-qc/1710.05832]}}.
\newblock
  doi:{\changeurlcolor{black}\href{https://doi.org/10.1103/PhysRevLett.119.161101}{\detokenize{10.1103/PhysRevLett.119.161101}}}.

\bibitem[Abbott \em{et~al.}(2017{\natexlab{b}})Abbott et~al.]{GBM:2017lvd}
Abbott, B.P.; others.
\newblock {Multi-messenger Observations of a Binary Neutron Star Merger}.
\newblock {\em Astrophys. J.} {\bf 2017}, {\em 848},~L12,
  \href{http://xxx.lanl.gov/abs/1710.05833}{{\normalfont
  [arXiv:astro-ph.HE/1710.05833]}}.
\newblock
  doi:{\changeurlcolor{black}\href{https://doi.org/10.3847/2041-8213/aa91c9}{\detokenize{10.3847/2041-8213/aa91c9}}}.

\bibitem[Abbott \em{et~al.}(2017{\natexlab{c}})Abbott et~al.]{Monitor:2017mdv}
Abbott, B.P.; others.
\newblock {Gravitational Waves and Gamma-rays from a Binary Neutron Star
  Merger: GW170817 and GRB 170817A}.
\newblock {\em Astrophys. J.} {\bf 2017}, {\em 848},~L13,
  \href{http://xxx.lanl.gov/abs/1710.05834}{{\normalfont
  [arXiv:astro-ph.HE/1710.05834]}}.
\newblock
  doi:{\changeurlcolor{black}\href{https://doi.org/10.3847/2041-8213/aa920c}{\detokenize{10.3847/2041-8213/aa920c}}}.

\bibitem[Dai \em{et~al.}(2018)Dai, Venumadhav, and Zackay]{Dai:2018dca}
Dai, L.; Venumadhav, T.; Zackay, B.
\newblock {Parameter Estimation for GW170817 using Relative Binning} {\bf
  2018}.
\newblock  \href{http://xxx.lanl.gov/abs/1806.08793}{{\normalfont
  [arXiv:gr-qc/1806.08793]}}.

\bibitem[De \em{et~al.}(2018)De, Finstad, Lattimer, Brown, Berger, and
  Biwer]{De:2018uhw}
De, S.; Finstad, D.; Lattimer, J.M.; Brown, D.A.; Berger, E.; Biwer, C.M.
\newblock {Tidal Deformabilities and Radii of Neutron Stars from the
  Observation of GW170817}.
\newblock {\em Phys. Rev. Lett.} {\bf 2018}, {\em 121},~091102,
  \href{http://xxx.lanl.gov/abs/1804.08583}{{\normalfont
  [arXiv:astro-ph.HE/1804.08583]}}.
\newblock [Erratum: Phys. Rev. Lett.121,no.25,259902(2018)],
  doi:{\changeurlcolor{black}\href{https://doi.org/10.1103/PhysRevLett.121.259902,
  10.1103/PhysRevLett.121.091102}{\detokenize{10.1103/PhysRevLett.121.259902,
  10.1103/PhysRevLett.121.091102}}}.

\bibitem[Abbott \em{et~al.}(2019)Abbott et~al.]{Abbott:2018wiz}
Abbott, B.P.; others.
\newblock {Properties of the binary neutron star merger GW170817}.
\newblock {\em Phys. Rev.} {\bf 2019}, {\em X9},~011001,
  \href{http://xxx.lanl.gov/abs/1805.11579}{{\normalfont
  [arXiv:gr-qc/1805.11579]}}.
\newblock
  doi:{\changeurlcolor{black}\href{https://doi.org/10.1103/PhysRevX.9.011001}{\detokenize{10.1103/PhysRevX.9.011001}}}.

\bibitem[Abbott \em{et~al.}(2018{\natexlab{a}})Abbott et~al.]{Abbott:2018exr}
Abbott, B.P.; others.
\newblock {GW170817: Measurements of neutron star radii and equation of state}.
\newblock {\em Phys. Rev. Lett.} {\bf 2018}, {\em 121},~161101,
  \href{http://xxx.lanl.gov/abs/1805.11581}{{\normalfont
  [arXiv:gr-qc/1805.11581]}}.
\newblock
  doi:{\changeurlcolor{black}\href{https://doi.org/10.1103/PhysRevLett.121.161101}{\detokenize{10.1103/PhysRevLett.121.161101}}}.

\bibitem[Abbott \em{et~al.}(2018{\natexlab{b}})Abbott
  et~al.]{LIGOScientific:2018mvr}
Abbott, B.P.; others.
\newblock {GWTC-1: A Gravitational-Wave Transient Catalog of Compact Binary
  Mergers Observed by LIGO and Virgo during the First and Second Observing
  Runs} {\bf 2018}.
\newblock  \href{http://xxx.lanl.gov/abs/1811.12907}{{\normalfont
  [arXiv:astro-ph.HE/1811.12907]}}.

\bibitem[Radice \em{et~al.}(2018)Radice, Perego, Zappa, and
  Bernuzzi]{Radice:2017lry}
Radice, D.; Perego, A.; Zappa, F.; Bernuzzi, S.
\newblock {GW170817: Joint Constraint on the Neutron Star Equation of State
  from Multimessenger Observations}.
\newblock {\em Astrophys. J.} {\bf 2018}, {\em 852},~L29,
  \href{http://xxx.lanl.gov/abs/1711.03647}{{\normalfont
  [arXiv:astro-ph.HE/1711.03647]}}.
\newblock
  doi:{\changeurlcolor{black}\href{https://doi.org/10.3847/2041-8213/aaa402}{\detokenize{10.3847/2041-8213/aaa402}}}.

\bibitem[Bauswein \em{et~al.}(2017)Bauswein, Just, Janka, and
  Stergioulas]{Bauswein:2017vtn}
Bauswein, A.; Just, O.; Janka, H.T.; Stergioulas, N.
\newblock {Neutron-star radius constraints from GW170817 and future
  detections}.
\newblock {\em Astrophys. J.} {\bf 2017}, {\em 850},~L34,
  \href{http://xxx.lanl.gov/abs/1710.06843}{{\normalfont
  [arXiv:astro-ph.HE/1710.06843]}}.
\newblock
  doi:{\changeurlcolor{black}\href{https://doi.org/10.3847/2041-8213/aa9994}{\detokenize{10.3847/2041-8213/aa9994}}}.

\bibitem[Coughlin \em{et~al.}(2018)Coughlin et~al.]{Coughlin:2018miv}
Coughlin, M.W.; others.
\newblock {Constraints on the neutron star equation of state from AT2017gfo
  using radiative transfer simulations} {\bf 2018}.
\newblock  \href{http://xxx.lanl.gov/abs/1805.09371}{{\normalfont
  [arXiv:astro-ph.HE/1805.09371]}}.
\newblock
  doi:{\changeurlcolor{black}\href{https://doi.org/10.1093/mnras/sty2174}{\detokenize{10.1093/mnras/sty2174}}}.

\bibitem[Radice and Dai(2018)]{Radice:2018ozg}
Radice, D.; Dai, L.
\newblock {Multimessenger Parameter Estimation of GW170817} {\bf 2018}.
\newblock  \href{http://xxx.lanl.gov/abs/1810.12917}{{\normalfont
  [arXiv:astro-ph.HE/1810.12917]}}.

\bibitem[Coughlin \em{et~al.}(2018)Coughlin, Dietrich, Margalit, and
  Metzger]{Coughlin:2018fis}
Coughlin, M.W.; Dietrich, T.; Margalit, B.; Metzger, B.D.
\newblock {Multi-messenger Bayesian parameter inference of a binary
  neutron-star merger} {\bf 2018}.
\newblock  \href{http://xxx.lanl.gov/abs/1812.04803}{{\normalfont
  [arXiv:astro-ph.HE/1812.04803]}}.

\bibitem[Annala \em{et~al.}(2018)Annala, Gorda, Kurkela, and
  Vuorinen]{Annala:2017llu}
Annala, E.; Gorda, T.; Kurkela, A.; Vuorinen, A.
\newblock {Gravitational-wave constraints on the neutron-star-matter Equation
  of State}.
\newblock {\em Phys. Rev. Lett.} {\bf 2018}, {\em 120},~172703,
  \href{http://xxx.lanl.gov/abs/1711.02644}{{\normalfont
  [arXiv:astro-ph.HE/1711.02644]}}.
\newblock
  doi:{\changeurlcolor{black}\href{https://doi.org/10.1103/PhysRevLett.120.172703}{\detokenize{10.1103/PhysRevLett.120.172703}}}.

\bibitem[Most \em{et~al.}(2018)Most, Weih, Rezzolla, and
  Schaffner-Bielich]{Most:2018hfd}
Most, E.R.; Weih, L.R.; Rezzolla, L.; Schaffner-Bielich, J.
\newblock {New constraints on radii and tidal deformabilities of neutron stars
  from GW170817}.
\newblock {\em Phys. Rev. Lett.} {\bf 2018}, {\em 120},~261103,
  \href{http://xxx.lanl.gov/abs/1803.00549}{{\normalfont
  [arXiv:gr-qc/1803.00549]}}.
\newblock
  doi:{\changeurlcolor{black}\href{https://doi.org/10.1103/PhysRevLett.120.261103}{\detokenize{10.1103/PhysRevLett.120.261103}}}.

\bibitem[Annala \em{et~al.}(2019)Annala, Gorda, Kurkela, Nättilä, and
  Vuorinen]{Annala:2019puf}
Annala, E.; Gorda, T.; Kurkela, A.; Nättilä, J.; Vuorinen, A.
\newblock {Quark-matter cores in neutron stars} {\bf 2019}.
\newblock  \href{http://xxx.lanl.gov/abs/1903.09121}{{\normalfont
  [arXiv:astro-ph.HE/1903.09121]}}.

\bibitem[Most \em{et~al.}(2019)Most, Papenfort, Dexheimer, Hanauske, Schramm,
  Stöcker, and Rezzolla]{Most:2018eaw}
Most, E.R.; Papenfort, L.J.; Dexheimer, V.; Hanauske, M.; Schramm, S.;
  Stöcker, H.; Rezzolla, L.
\newblock {Signatures of quark-hadron phase transitions in general-relativistic
  neutron-star mergers}.
\newblock {\em Phys. Rev. Lett.} {\bf 2019}, {\em 122},~061101,
  \href{http://xxx.lanl.gov/abs/1807.03684}{{\normalfont
  [arXiv:astro-ph.HE/1807.03684]}}.
\newblock
  doi:{\changeurlcolor{black}\href{https://doi.org/10.1103/PhysRevLett.122.061101}{\detokenize{10.1103/PhysRevLett.122.061101}}}.

\bibitem[Bauswein \em{et~al.}(2019)Bauswein, Bastian, Blaschke, Chatziioannou,
  Clark, Fischer, and Oertel]{Bauswein:2018bma}
Bauswein, A.; Bastian, N.U.F.; Blaschke, D.B.; Chatziioannou, K.; Clark, J.A.;
  Fischer, T.; Oertel, M.
\newblock {Identifying a first-order phase transition in neutron star mergers
  through gravitational waves}.
\newblock {\em Phys. Rev. Lett.} {\bf 2019}, {\em 122},~061102,
  \href{http://xxx.lanl.gov/abs/1809.01116}{{\normalfont
  [arXiv:astro-ph.HE/1809.01116]}}.
\newblock
  doi:{\changeurlcolor{black}\href{https://doi.org/10.1103/PhysRevLett.122.061102}{\detokenize{10.1103/PhysRevLett.122.061102}}}.

\bibitem[Radice \em{et~al.}(2014)Radice, Rezzolla, and
  Galeazzi]{Radice:2013hxh}
Radice, D.; Rezzolla, L.; Galeazzi, F.
\newblock {Beyond second-order convergence in simulations of binary neutron
  stars in full general-relativity}.
\newblock {\em Mon. Not. Roy. Astron. Soc.} {\bf 2014}, {\em 437},~L46--L50,
  \href{http://xxx.lanl.gov/abs/1306.6052}{{\normalfont
  [arXiv:gr-qc/1306.6052]}}.
\newblock
  doi:{\changeurlcolor{black}\href{https://doi.org/10.1093/mnrasl/slt137}{\detokenize{10.1093/mnrasl/slt137}}}.

\bibitem[Hotokezaka \em{et~al.}(2015)Hotokezaka, Kyutoku, Okawa, and
  Shibata]{Hotokezaka:2015xka}
Hotokezaka, K.; Kyutoku, K.; Okawa, H.; Shibata, M.
\newblock {Exploring tidal effects of coalescing binary neutron stars in
  numerical relativity. II. Long-term simulations}.
\newblock {\em Phys. Rev.} {\bf 2015}, {\em D91},~064060,
  \href{http://xxx.lanl.gov/abs/1502.03457}{{\normalfont
  [arXiv:gr-qc/1502.03457]}}.
\newblock
  doi:{\changeurlcolor{black}\href{https://doi.org/10.1103/PhysRevD.91.064060}{\detokenize{10.1103/PhysRevD.91.064060}}}.

\bibitem[Dietrich \em{et~al.}(2017)Dietrich, Bernuzzi, and
  Tichy]{Dietrich:2017aum}
Dietrich, T.; Bernuzzi, S.; Tichy, W.
\newblock {Closed-form tidal approximants for binary neutron star gravitational
  waveforms constructed from high-resolution numerical relativity simulations}.
\newblock {\em Phys. Rev.} {\bf 2017}, {\em D96},~121501,
  \href{http://xxx.lanl.gov/abs/1706.02969}{{\normalfont
  [arXiv:gr-qc/1706.02969]}}.
\newblock
  doi:{\changeurlcolor{black}\href{https://doi.org/10.1103/PhysRevD.96.121501}{\detokenize{10.1103/PhysRevD.96.121501}}}.

\bibitem[Kiuchi \em{et~al.}(2017)Kiuchi, Kawaguchi, Kyutoku, Sekiguchi,
  Shibata, and Taniguchi]{Kiuchi:2017pte}
Kiuchi, K.; Kawaguchi, K.; Kyutoku, K.; Sekiguchi, Y.; Shibata, M.; Taniguchi,
  K.
\newblock {Sub-radian-accuracy gravitational waveforms of coalescing binary
  neutron stars in numerical relativity}.
\newblock {\em Phys. Rev.} {\bf 2017}, {\em D96},~084060,
  \href{http://xxx.lanl.gov/abs/1708.08926}{{\normalfont
  [arXiv:astro-ph.HE/1708.08926]}}.
\newblock
  doi:{\changeurlcolor{black}\href{https://doi.org/10.1103/PhysRevD.96.084060}{\detokenize{10.1103/PhysRevD.96.084060}}}.

\bibitem[Dietrich \em{et~al.}(2018)Dietrich, Radice, Bernuzzi, Zappa, Perego,
  Brügmann, Chaurasia, Dudi, Tichy, and Ujevic]{Dietrich:2018phi}
Dietrich, T.; Radice, D.; Bernuzzi, S.; Zappa, F.; Perego, A.; Brügmann, B.;
  Chaurasia, S.V.; Dudi, R.; Tichy, W.; Ujevic, M.
\newblock {CoRe database of binary neutron star merger waveforms}.
\newblock {\em Class. Quant. Grav.} {\bf 2018}, {\em 35},~24LT01,
  \href{http://xxx.lanl.gov/abs/1806.01625}{{\normalfont
  [arXiv:gr-qc/1806.01625]}}.
\newblock
  doi:{\changeurlcolor{black}\href{https://doi.org/10.1088/1361-6382/aaebc0}{\detokenize{10.1088/1361-6382/aaebc0}}}.

\bibitem[Rezzolla \em{et~al.}(2011)Rezzolla, Giacomazzo, Baiotti, Granot,
  Kouveliotou, and Aloy]{Rezzolla:2011da}
Rezzolla, L.; Giacomazzo, B.; Baiotti, L.; Granot, J.; Kouveliotou, C.; Aloy,
  M.A.
\newblock {The missing link: Merging neutron stars naturally produce jet-like
  structures and can power short Gamma-Ray Bursts}.
\newblock {\em Astrophys. J.} {\bf 2011}, {\em 732},~L6,
  \href{http://xxx.lanl.gov/abs/1101.4298}{{\normalfont
  [arXiv:astro-ph.HE/1101.4298]}}.
\newblock
  doi:{\changeurlcolor{black}\href{https://doi.org/10.1088/2041-8205/732/1/L6}{\detokenize{10.1088/2041-8205/732/1/L6}}}.

\bibitem[Neilsen \em{et~al.}(2014)Neilsen, Liebling, Anderson, Lehner,
  O'Connor, and Palenzuela]{Neilsen:2014hha}
Neilsen, D.; Liebling, S.L.; Anderson, M.; Lehner, L.; O'Connor, E.;
  Palenzuela, C.
\newblock {Magnetized Neutron Stars With Realistic Equations of State and
  Neutrino Cooling}.
\newblock {\em Phys. Rev.} {\bf 2014}, {\em D89},~104029,
  \href{http://xxx.lanl.gov/abs/1403.3680}{{\normalfont
  [arXiv:gr-qc/1403.3680]}}.
\newblock
  doi:{\changeurlcolor{black}\href{https://doi.org/10.1103/PhysRevD.89.104029}{\detokenize{10.1103/PhysRevD.89.104029}}}.

\bibitem[Sekiguchi \em{et~al.}(2015)Sekiguchi, Kiuchi, Kyutoku, and
  Shibata]{Sekiguchi:2015dma}
Sekiguchi, Y.; Kiuchi, K.; Kyutoku, K.; Shibata, M.
\newblock {Dynamical mass ejection from binary neutron star mergers:
  Radiation-hydrodynamics study in general relativity}.
\newblock {\em Phys. Rev.} {\bf 2015}, {\em D91},~064059,
  \href{http://xxx.lanl.gov/abs/1502.06660}{{\normalfont
  [arXiv:astro-ph.HE/1502.06660]}}.
\newblock
  doi:{\changeurlcolor{black}\href{https://doi.org/10.1103/PhysRevD.91.064059}{\detokenize{10.1103/PhysRevD.91.064059}}}.

\bibitem[Palenzuela \em{et~al.}(2015)Palenzuela, Liebling, Neilsen, Lehner,
  Caballero, O'Connor, and Anderson]{Palenzuela:2015dqa}
Palenzuela, C.; Liebling, S.L.; Neilsen, D.; Lehner, L.; Caballero, O.L.;
  O'Connor, E.; Anderson, M.
\newblock {Effects of the microphysical Equation of State in the mergers of
  magnetized Neutron Stars With Neutrino Cooling}.
\newblock {\em Phys. Rev.} {\bf 2015}, {\em D92},~044045,
  \href{http://xxx.lanl.gov/abs/1505.01607}{{\normalfont
  [arXiv:gr-qc/1505.01607]}}.
\newblock
  doi:{\changeurlcolor{black}\href{https://doi.org/10.1103/PhysRevD.92.044045}{\detokenize{10.1103/PhysRevD.92.044045}}}.

\bibitem[Foucart(2018)]{Foucart:2017mbt}
Foucart, F.
\newblock {Monte Carlo closure for moment-based transport schemes in general
  relativistic radiation hydrodynamic simulations}.
\newblock {\em Mon. Not. Roy. Astron. Soc.} {\bf 2018}, {\em 475},~4186--4207,
  \href{http://xxx.lanl.gov/abs/1708.08452}{{\normalfont
  [arXiv:astro-ph.HE/1708.08452]}}.
\newblock
  doi:{\changeurlcolor{black}\href{https://doi.org/10.1093/mnras/sty108}{\detokenize{10.1093/mnras/sty108}}}.

\bibitem[Ruiz \em{et~al.}(2018)Ruiz, Shapiro, and Tsokaros]{Ruiz:2017due}
Ruiz, M.; Shapiro, S.L.; Tsokaros, A.
\newblock {GW170817, General Relativistic Magnetohydrodynamic Simulations, and
  the Neutron Star Maximum Mass}.
\newblock {\em Phys. Rev.} {\bf 2018}, {\em D97},~021501,
  \href{http://xxx.lanl.gov/abs/1711.00473}{{\normalfont
  [arXiv:astro-ph.HE/1711.00473]}}.
\newblock
  doi:{\changeurlcolor{black}\href{https://doi.org/10.1103/PhysRevD.97.021501}{\detokenize{10.1103/PhysRevD.97.021501}}}.

\bibitem[Ciolfi \em{et~al.}(2017)Ciolfi, Kastaun, Giacomazzo, Endrizzi, Siegel,
  and Perna]{Ciolfi:2017uak}
Ciolfi, R.; Kastaun, W.; Giacomazzo, B.; Endrizzi, A.; Siegel, D.M.; Perna, R.
\newblock {General relativistic magnetohydrodynamic simulations of binary
  neutron star mergers forming a long-lived neutron star}.
\newblock {\em Phys. Rev.} {\bf 2017}, {\em D95},~063016,
  \href{http://xxx.lanl.gov/abs/1701.08738}{{\normalfont
  [arXiv:astro-ph.HE/1701.08738]}}.
\newblock
  doi:{\changeurlcolor{black}\href{https://doi.org/10.1103/PhysRevD.95.063016}{\detokenize{10.1103/PhysRevD.95.063016}}}.

\bibitem[Kiuchi \em{et~al.}(2018)Kiuchi, Kyutoku, Sekiguchi, and
  Shibata]{Kiuchi:2017zzg}
Kiuchi, K.; Kyutoku, K.; Sekiguchi, Y.; Shibata, M.
\newblock {Global simulations of strongly magnetized remnant massive neutron
  stars formed in binary neutron star mergers}.
\newblock {\em Phys. Rev.} {\bf 2018}, {\em D97},~124039,
  \href{http://xxx.lanl.gov/abs/1710.01311}{{\normalfont
  [arXiv:astro-ph.HE/1710.01311]}}.
\newblock
  doi:{\changeurlcolor{black}\href{https://doi.org/10.1103/PhysRevD.97.124039}{\detokenize{10.1103/PhysRevD.97.124039}}}.

\bibitem[Radice(2017)]{Radice:2017zta}
Radice, D.
\newblock {General-Relativistic Large-Eddy Simulations of Binary Neutron Star
  Mergers}.
\newblock {\em Astrophys. J.} {\bf 2017}, {\em 838},~L2,
  \href{http://xxx.lanl.gov/abs/1703.02046}{{\normalfont
  [arXiv:astro-ph.HE/1703.02046]}}.
\newblock
  doi:{\changeurlcolor{black}\href{https://doi.org/10.3847/2041-8213/aa6483}{\detokenize{10.3847/2041-8213/aa6483}}}.

\bibitem[Shibata \em{et~al.}(2017)Shibata, Kiuchi, and
  Sekiguchi]{Shibata:2017jyf}
Shibata, M.; Kiuchi, K.; Sekiguchi, Y.i.
\newblock {General relativistic viscous hydrodynamics of differentially
  rotating neutron stars}.
\newblock {\em Phys. Rev.} {\bf 2017}, {\em D95},~083005,
  \href{http://xxx.lanl.gov/abs/1703.10303}{{\normalfont
  [arXiv:astro-ph.HE/1703.10303]}}.
\newblock
  doi:{\changeurlcolor{black}\href{https://doi.org/10.1103/PhysRevD.95.083005}{\detokenize{10.1103/PhysRevD.95.083005}}}.

\bibitem[Pietri \em{et~al.}(2019)Pietri, Drago, Feo, Pagliara, Pasquali,
  Traversi, and Wiktorowicz]{Pietri:2019eb}
Pietri, R.d.; Drago, A.; Feo, A.; Pagliara, G.; Pasquali, M.; Traversi, S.;
  Wiktorowicz, G.
\newblock {Merger of Compact Stars in the Two-Families Scenario} {\bf 2019}.
\newblock  \href{http://xxx.lanl.gov/abs/1904.01545}{{\normalfont
  [arXiv:astro-ph.HE/1904.01545]}}.

\bibitem[Farhi and Jaffe(1984)]{Farhi:1984qu}
Farhi, E.; Jaffe, R.L.
\newblock {Strange Matter}.
\newblock {\em Phys. Rev.} {\bf 1984}, {\em D30},~2379.
\newblock
  doi:{\changeurlcolor{black}\href{https://doi.org/10.1103/PhysRevD.30.2379}{\detokenize{10.1103/PhysRevD.30.2379}}}.

\bibitem[Chodos \em{et~al.}(1974)Chodos, Jaffe, Johnson, Thorn, and
  Weisskopf]{Chodos:1974je}
Chodos, A.; Jaffe, R.L.; Johnson, K.; Thorn, C.B.; Weisskopf, V.F.
\newblock {A New Extended Model of Hadrons}.
\newblock {\em Phys. Rev.} {\bf 1974}, {\em D9},~3471--3495.
\newblock
  doi:{\changeurlcolor{black}\href{https://doi.org/10.1103/PhysRevD.9.3471}{\detokenize{10.1103/PhysRevD.9.3471}}}.

\bibitem[Detar(1979)]{DeTar:1979vb}
Detar, C.E.
\newblock {THE MIT BAG MODEL}.
\newblock  {NATO Advanced Study Institute in Elementary Particles: Quantum
  Flavordynamics, Quantum Chromodynamics and Unified Theories Boulder, Colo.,
  July 9-27, 1979},  1979, p. 393.

\bibitem[Alford \em{et~al.}(2005)Alford, Braby, Paris, and
  Reddy]{Alford:2004pf}
Alford, M.; Braby, M.; Paris, M.W.; Reddy, S.
\newblock {Hybrid stars that masquerade as neutron stars}.
\newblock {\em Astrophys. J.} {\bf 2005}, {\em 629},~969--978,
  \href{http://xxx.lanl.gov/abs/nucl-th/0411016}{{\normalfont
  [arXiv:nucl-th/nucl-th/0411016]}}.
\newblock
  doi:{\changeurlcolor{black}\href{https://doi.org/10.1086/430902}{\detokenize{10.1086/430902}}}.

\bibitem[Alford \em{et~al.}(2008)Alford, Schmitt, Rajagopal, and
  Schäfer]{Alford:2007xm}
Alford, M.G.; Schmitt, A.; Rajagopal, K.; Schäfer, T.
\newblock {Color superconductivity in dense quark matter}.
\newblock {\em Rev. Mod. Phys.} {\bf 2008}, {\em 80},~1455--1515,
  \href{http://xxx.lanl.gov/abs/0709.4635}{{\normalfont
  [arXiv:hep-ph/0709.4635]}}.
\newblock
  doi:{\changeurlcolor{black}\href{https://doi.org/10.1103/RevModPhys.80.1455}{\detokenize{10.1103/RevModPhys.80.1455}}}.

\bibitem[Lugones and Horvath(2002)]{Lugones:2002va}
Lugones, G.; Horvath, J.E.
\newblock {Color flavor locked strange matter}.
\newblock {\em Phys. Rev.} {\bf 2002}, {\em D66},~074017,
  \href{http://xxx.lanl.gov/abs/hep-ph/0211070}{{\normalfont
  [arXiv:hep-ph/hep-ph/0211070]}}.
\newblock
  doi:{\changeurlcolor{black}\href{https://doi.org/10.1103/PhysRevD.66.074017}{\detokenize{10.1103/PhysRevD.66.074017}}}.

\bibitem[Fraga \em{et~al.}(2001)Fraga, Pisarski, and
  Schaffner-Bielich]{Fraga:2001id}
Fraga, E.S.; Pisarski, R.D.; Schaffner-Bielich, J.
\newblock {Small, dense quark stars from perturbative QCD}.
\newblock {\em Phys. Rev.} {\bf 2001}, {\em D63},~121702,
  \href{http://xxx.lanl.gov/abs/hep-ph/0101143}{{\normalfont
  [arXiv:hep-ph/hep-ph/0101143]}}.
\newblock
  doi:{\changeurlcolor{black}\href{https://doi.org/10.1103/PhysRevD.63.121702}{\detokenize{10.1103/PhysRevD.63.121702}}}.

\bibitem[Klahn and Fischer(2015)]{Klahn:2015mfa}
Klahn, T.; Fischer, T.
\newblock {Vector interaction enhanced bag model for astrophysical
  applications}.
\newblock {\em Astrophys. J.} {\bf 2015}, {\em 810},~134,
  \href{http://xxx.lanl.gov/abs/1503.07442}{{\normalfont
  [arXiv:nucl-th/1503.07442]}}.
\newblock
  doi:{\changeurlcolor{black}\href{https://doi.org/10.1088/0004-637X/810/2/134}{\detokenize{10.1088/0004-637X/810/2/134}}}.

\bibitem[Pereira \em{et~al.}(2018)Pereira, Flores, and
  Lugones]{Pereira:2017rmp}
Pereira, J.P.; Flores, C.V.; Lugones, G.
\newblock {Phase transition effects on the dynamical stability of hybrid
  neutron stars}.
\newblock {\em Astrophys. J.} {\bf 2018}, {\em 860},~12,
  \href{http://xxx.lanl.gov/abs/1706.09371}{{\normalfont
  [arXiv:gr-qc/1706.09371]}}.
\newblock
  doi:{\changeurlcolor{black}\href{https://doi.org/10.3847/1538-4357/aabfbf}{\detokenize{10.3847/1538-4357/aabfbf}}}.

\bibitem[Bodmer(1971)]{Bodmer:1971we}
Bodmer, A.R.
\newblock {Collapsed nuclei}.
\newblock {\em Phys. Rev.} {\bf 1971}, {\em D4},~1601--1606.
\newblock
  doi:{\changeurlcolor{black}\href{https://doi.org/10.1103/PhysRevD.4.1601}{\detokenize{10.1103/PhysRevD.4.1601}}}.

\bibitem[Witten(1984)]{Witten:1984rs}
Witten, E.
\newblock {Cosmic Separation of Phases}.
\newblock {\em Phys. Rev.} {\bf 1984}, {\em D30},~272--285.
\newblock
  doi:{\changeurlcolor{black}\href{https://doi.org/10.1103/PhysRevD.30.272}{\detokenize{10.1103/PhysRevD.30.272}}}.

\bibitem[Glendenning(2000)]{Glendenning_book}
Glendenning, N.K.
\newblock {\em Compact Stars: Nuclear Physics, Particle Physics, and General
  Relativity}, 2 ed.; Springer: The address,  2000.

\bibitem[Lugones \em{et~al.}(2013)Lugones, Grunfeld, and
  Al~Ajmi]{Lugones:2013ema}
Lugones, G.; Grunfeld, A.G.; Al~Ajmi, M.
\newblock {Surface tension and curvature energy of quark matter in the
  Nambu-Jona-Lasinio model}.
\newblock {\em Phys. Rev.} {\bf 2013}, {\em C88},~045803,
  \href{http://xxx.lanl.gov/abs/1308.1452}{{\normalfont
  [arXiv:hep-ph/1308.1452]}}.
\newblock
  doi:{\changeurlcolor{black}\href{https://doi.org/10.1103/PhysRevC.88.045803}{\detokenize{10.1103/PhysRevC.88.045803}}}.

\bibitem[Nambu and Jona-Lasinio(1961)]{Nambu:1961tp}
Nambu, Y.; Jona-Lasinio, G.
\newblock {Dynamical Model of Elementary Particles Based on an Analogy with
  Superconductivity. 1.}
\newblock {\em Phys. Rev.} {\bf 1961}, {\em 122},~345--358.
\newblock [,127(1961)],
  doi:{\changeurlcolor{black}\href{https://doi.org/10.1103/PhysRev.122.345}{\detokenize{10.1103/PhysRev.122.345}}}.

\bibitem[Douchin and Haensel(2001)]{Douchin:2001sv}
Douchin, F.; Haensel, P.
\newblock {A unified equation of state of dense matter and neutron star
  structure}.
\newblock {\em Astron. Astrophys.} {\bf 2001}, {\em 380},~151,
  \href{http://xxx.lanl.gov/abs/astro-ph/0111092}{{\normalfont
  [arXiv:astro-ph/astro-ph/0111092]}}.
\newblock
  doi:{\changeurlcolor{black}\href{https://doi.org/10.1051/0004-6361:20011402}{\detokenize{10.1051/0004-6361:20011402}}}.

\bibitem[Bruegmann \em{et~al.}(2008)Bruegmann, Gonzalez, Hannam, Husa,
  Sperhake, and Tichy]{Bruegmann:2006at}
Bruegmann, B.; Gonzalez, J.A.; Hannam, M.; Husa, S.; Sperhake, U.; Tichy, W.
\newblock {Calibration of Moving Puncture Simulations}.
\newblock {\em Phys. Rev.} {\bf 2008}, {\em D77},~024027,
  \href{http://xxx.lanl.gov/abs/gr-qc/0610128}{{\normalfont
  [arXiv:gr-qc/gr-qc/0610128]}}.
\newblock
  doi:{\changeurlcolor{black}\href{https://doi.org/10.1103/PhysRevD.77.024027}{\detokenize{10.1103/PhysRevD.77.024027}}}.

\bibitem[Thierfelder \em{et~al.}(2011)Thierfelder, Bernuzzi, and
  Bruegmann]{Thierfelder:2011yi}
Thierfelder, M.; Bernuzzi, S.; Bruegmann, B.
\newblock {Numerical relativity simulations of binary neutron stars}.
\newblock {\em Phys. Rev.} {\bf 2011}, {\em D84},~044012,
  \href{http://xxx.lanl.gov/abs/1104.4751}{{\normalfont
  [arXiv:gr-qc/1104.4751]}}.
\newblock
  doi:{\changeurlcolor{black}\href{https://doi.org/10.1103/PhysRevD.84.044012}{\detokenize{10.1103/PhysRevD.84.044012}}}.

\bibitem[Dietrich \em{et~al.}(2015)Dietrich, Bernuzzi, Ujevic, and
  Brügmann]{Dietrich:2015iva}
Dietrich, T.; Bernuzzi, S.; Ujevic, M.; Brügmann, B.
\newblock {Numerical relativity simulations of neutron star merger remnants
  using conservative mesh refinement}.
\newblock {\em Phys. Rev.} {\bf 2015}, {\em D91},~124041,
  \href{http://xxx.lanl.gov/abs/1504.01266}{{\normalfont
  [arXiv:gr-qc/1504.01266]}}.
\newblock
  doi:{\changeurlcolor{black}\href{https://doi.org/10.1103/PhysRevD.91.124041}{\detokenize{10.1103/PhysRevD.91.124041}}}.

\bibitem[Bernuzzi and Dietrich(2016)]{Bernuzzi:2016pie}
Bernuzzi, S.; Dietrich, T.
\newblock {Gravitational waveforms from binary neutron star mergers with
  high-order weighted-essentially-nonoscillatory schemes in numerical
  relativity}.
\newblock {\em Phys. Rev.} {\bf 2016}, {\em D94},~064062,
  \href{http://xxx.lanl.gov/abs/1604.07999}{{\normalfont
  [arXiv:gr-qc/1604.07999]}}.
\newblock
  doi:{\changeurlcolor{black}\href{https://doi.org/10.1103/PhysRevD.94.064062}{\detokenize{10.1103/PhysRevD.94.064062}}}.

\bibitem[Tichy(2009)]{Tichy:2009yr}
Tichy, W.
\newblock {A New numerical method to construct binary neutron star initial
  data}.
\newblock {\em Class. Quant. Grav.} {\bf 2009}, {\em 26},~175018,
  \href{http://xxx.lanl.gov/abs/0908.0620}{{\normalfont
  [arXiv:gr-qc/0908.0620]}}.
\newblock
  doi:{\changeurlcolor{black}\href{https://doi.org/10.1088/0264-9381/26/17/175018}{\detokenize{10.1088/0264-9381/26/17/175018}}}.

\bibitem[Read \em{et~al.}(2009)Read, Lackey, Owen, and Friedman]{Read:2008iy}
Read, J.S.; Lackey, B.D.; Owen, B.J.; Friedman, J.L.
\newblock {Constraints on a phenomenologically parameterized neutron-star
  equation of state}.
\newblock {\em Phys. Rev.} {\bf 2009}, {\em D79},~124032,
  \href{http://xxx.lanl.gov/abs/0812.2163}{{\normalfont
  [arXiv:astro-ph/0812.2163]}}.
\newblock
  doi:{\changeurlcolor{black}\href{https://doi.org/10.1103/PhysRevD.79.124032}{\detokenize{10.1103/PhysRevD.79.124032}}}.

\bibitem[Wilson and Mathews(1995)]{Wilson:1995uh}
Wilson, J.R.; Mathews, G.J.
\newblock {Instabilities in Close Neutron Star Binaries}.
\newblock {\em Phys. Rev. Lett.} {\bf 1995}, {\em 75},~4161--4164.
\newblock
  doi:{\changeurlcolor{black}\href{https://doi.org/10.1103/PhysRevLett.75.4161}{\detokenize{10.1103/PhysRevLett.75.4161}}}.

\bibitem[Wilson \em{et~al.}(1996)Wilson, Mathews, and
  Marronetti]{Wilson:1996ty}
Wilson, J.R.; Mathews, G.J.; Marronetti, P.
\newblock {Relativistic numerical model for close neutron star binaries}.
\newblock {\em Phys. Rev.} {\bf 1996}, {\em D54},~1317--1331,
  \href{http://xxx.lanl.gov/abs/gr-qc/9601017}{{\normalfont
  [arXiv:gr-qc/gr-qc/9601017]}}.
\newblock
  doi:{\changeurlcolor{black}\href{https://doi.org/10.1103/PhysRevD.54.1317}{\detokenize{10.1103/PhysRevD.54.1317}}}.

\bibitem[York(1999)]{York:1998hy}
York, Jr., J.W.
\newblock {Conformal 'thin sandwich' data for the initial-value problem}.
\newblock {\em Phys. Rev. Lett.} {\bf 1999}, {\em 82},~1350--1353,
  \href{http://xxx.lanl.gov/abs/gr-qc/9810051}{{\normalfont
  [arXiv:gr-qc/gr-qc/9810051]}}.
\newblock
  doi:{\changeurlcolor{black}\href{https://doi.org/10.1103/PhysRevLett.82.1350}{\detokenize{10.1103/PhysRevLett.82.1350}}}.

\bibitem[Tichy(2012)]{Tichy:2012rp}
Tichy, W.
\newblock {Constructing quasi-equilibrium initial data for binary neutron stars
  with arbitrary spins}.
\newblock {\em Phys. Rev. D} {\bf 2012}, {\em 86},~064024,
  \href{http://xxx.lanl.gov/abs/1209.5336}{{\normalfont
  [arXiv:gr-qc/1209.5336]}}.
\newblock
  doi:{\changeurlcolor{black}\href{https://doi.org/10.1103/PhysRevD.86.064024}{\detokenize{10.1103/PhysRevD.86.064024}}}.

\bibitem[Dietrich \em{et~al.}(2015)Dietrich, Moldenhauer, Johnson-McDaniel,
  Bernuzzi, Markakis, Brügmann, and Tichy]{Dietrich:2015pxa}
Dietrich, T.; Moldenhauer, N.; Johnson-McDaniel, N.K.; Bernuzzi, S.; Markakis,
  C.M.; Brügmann, B.; Tichy, W.
\newblock {Binary Neutron Stars with Generic Spin, Eccentricity, Mass ratio,
  and Compactness - Quasi-equilibrium Sequences and First Evolutions}.
\newblock {\em Phys. Rev.} {\bf 2015}, {\em D92},~124007,
  \href{http://xxx.lanl.gov/abs/1507.07100}{{\normalfont
  [arXiv:gr-qc/1507.07100]}}.
\newblock
  doi:{\changeurlcolor{black}\href{https://doi.org/10.1103/PhysRevD.92.124007}{\detokenize{10.1103/PhysRevD.92.124007}}}.

\bibitem[Moldenhauer \em{et~al.}(2014)Moldenhauer, Markakis, Johnson-McDaniel,
  Tichy, and Brügmann]{Moldenhauer:2014yaa}
Moldenhauer, N.; Markakis, C.M.; Johnson-McDaniel, N.K.; Tichy, W.; Brügmann,
  B.
\newblock {Initial data for binary neutron stars with adjustable eccentricity}.
\newblock {\em Phys. Rev.} {\bf 2014}, {\em D90},~084043,
  \href{http://xxx.lanl.gov/abs/1408.4136}{{\normalfont
  [arXiv:gr-qc/1408.4136]}}.
\newblock
  doi:{\changeurlcolor{black}\href{https://doi.org/10.1103/PhysRevD.90.084043}{\detokenize{10.1103/PhysRevD.90.084043}}}.

\bibitem[Kyutoku \em{et~al.}(2014)Kyutoku, Shibata, and
  Taniguchi]{Kyutoku:2014yba}
Kyutoku, K.; Shibata, M.; Taniguchi, K.
\newblock {Reducing orbital eccentricity in initial data of binary neutron
  stars}.
\newblock {\em Phys. Rev.} {\bf 2014}, {\em D90},~064006,
  \href{http://xxx.lanl.gov/abs/1405.6207}{{\normalfont
  [arXiv:gr-qc/1405.6207]}}.
\newblock
  doi:{\changeurlcolor{black}\href{https://doi.org/10.1103/PhysRevD.90.064006}{\detokenize{10.1103/PhysRevD.90.064006}}}.

\bibitem[Bernuzzi and Hilditch(2010)]{Bernuzzi:2009ex}
Bernuzzi, S.; Hilditch, D.
\newblock {Constraint violation in free evolution schemes: Comparing BSSNOK
  with a conformal decomposition of Z4}.
\newblock {\em Phys. Rev.} {\bf 2010}, {\em D81},~084003,
  \href{http://xxx.lanl.gov/abs/0912.2920}{{\normalfont
  [arXiv:gr-qc/0912.2920]}}.
\newblock
  doi:{\changeurlcolor{black}\href{https://doi.org/10.1103/PhysRevD.81.084003}{\detokenize{10.1103/PhysRevD.81.084003}}}.

\bibitem[Weyhausen \em{et~al.}(2012)Weyhausen, Bernuzzi, and
  Hilditch]{Weyhausen:2011cg}
Weyhausen, A.; Bernuzzi, S.; Hilditch, D.
\newblock {Constraint damping for the Z4c formulation of general relativity}.
\newblock {\em Phys. Rev.} {\bf 2012}, {\em D85},~024038,
  \href{http://xxx.lanl.gov/abs/1107.5539}{{\normalfont
  [arXiv:gr-qc/1107.5539]}}.
\newblock
  doi:{\changeurlcolor{black}\href{https://doi.org/10.1103/PhysRevD.85.024038}{\detokenize{10.1103/PhysRevD.85.024038}}}.

\bibitem[Hilditch \em{et~al.}(2013)Hilditch, Bernuzzi, Thierfelder, Cao, Tichy,
  and Bruegmann]{Hilditch:2012fp}
Hilditch, D.; Bernuzzi, S.; Thierfelder, M.; Cao, Z.; Tichy, W.; Bruegmann, B.
\newblock {Compact binary evolutions with the Z4c formulation}.
\newblock {\em Phys. Rev.} {\bf 2013}, {\em D88},~084057,
  \href{http://xxx.lanl.gov/abs/1212.2901}{{\normalfont
  [arXiv:gr-qc/1212.2901]}}.
\newblock
  doi:{\changeurlcolor{black}\href{https://doi.org/10.1103/PhysRevD.88.084057}{\detokenize{10.1103/PhysRevD.88.084057}}}.

\bibitem[Borges \em{et~al.}(2008)Borges, Carmona, Costa, and Don]{Borges:2008a}
Borges, R.; Carmona, M.; Costa, B.; Don, W.S.
\newblock An improved weighted essentially non-oscillatory scheme for
  hyperbolic conservation laws.
\newblock {\em Journal of Computational Physics} {\bf 2008}, {\em
  227},~3191--3211.
\newblock
  doi:{\changeurlcolor{black}\href{https://doi.org/10.1016/j.jcp.2007.11.038}{\detokenize{10.1016/j.jcp.2007.11.038}}}.

\bibitem[Dietrich \em{et~al.}(2018)Dietrich, Bernuzzi, Bruegmann, and
  Tichy]{Dietrich:2018upm}
Dietrich, T.; Bernuzzi, S.; Bruegmann, B.; Tichy, W.
\newblock {High-resolution numerical relativity simulations of spinning binary
  neutron star mergers}.
\newblock  {Proceedings, 26th Euromicro International Conference on Parallel,
  Distributed and Network-based Processing (PDP 2018): Cambridge, UK, March
  21-23, 2018},  2018, pp. 682--689,
  \href{http://xxx.lanl.gov/abs/1803.07965}{{\normalfont
  [arXiv:gr-qc/1803.07965]}}.
\newblock
  doi:{\changeurlcolor{black}\href{https://doi.org/10.1109/PDP2018.2018.00113}{\detokenize{10.1109/PDP2018.2018.00113}}}.

\bibitem[Faber and Rasio(2012)]{Faber:2012rw}
Faber, J.A.; Rasio, F.A.
\newblock {Binary Neutron Star Mergers}.
\newblock {\em Living Rev.Rel.} {\bf 2012}, {\em 15},~8,
  \href{http://xxx.lanl.gov/abs/1204.3858}{{\normalfont
  [arXiv:gr-qc/1204.3858]}}.

\bibitem[Dietrich \em{et~al.}(2019)Dietrich et~al.]{Dietrich:2018uni}
Dietrich, T.; others.
\newblock {Matter imprints in waveform models for neutron star binaries: Tidal
  and self-spin effects}.
\newblock {\em Phys. Rev.} {\bf 2019}, {\em D99},~024029,
  \href{http://xxx.lanl.gov/abs/1804.02235}{{\normalfont
  [arXiv:gr-qc/1804.02235]}}.
\newblock
  doi:{\changeurlcolor{black}\href{https://doi.org/10.1103/PhysRevD.99.024029}{\detokenize{10.1103/PhysRevD.99.024029}}}.

\bibitem[Dietrich \em{et~al.}(2017{\natexlab{a}})Dietrich, Ujevic, Tichy,
  Bernuzzi, and Bruegmann]{Dietrich:2016hky}
Dietrich, T.; Ujevic, M.; Tichy, W.; Bernuzzi, S.; Bruegmann, B.
\newblock {Gravitational waves and mass ejecta from binary neutron star
  mergers: Effect of the mass-ratio}.
\newblock {\em Phys. Rev.} {\bf 2017}, {\em D95},~024029,
  \href{http://xxx.lanl.gov/abs/1607.06636}{{\normalfont
  [arXiv:gr-qc/1607.06636]}}.
\newblock
  doi:{\changeurlcolor{black}\href{https://doi.org/10.1103/PhysRevD.95.024029}{\detokenize{10.1103/PhysRevD.95.024029}}}.

\bibitem[Dietrich \em{et~al.}(2017{\natexlab{b}})Dietrich, Bernuzzi, Ujevic,
  and Tichy]{Dietrich:2016lyp}
Dietrich, T.; Bernuzzi, S.; Ujevic, M.; Tichy, W.
\newblock {Gravitational waves and mass ejecta from binary neutron star
  mergers: Effect of the stars' rotation}.
\newblock {\em Phys. Rev.} {\bf 2017}, {\em D95},~044045,
  \href{http://xxx.lanl.gov/abs/1611.07367}{{\normalfont
  [arXiv:gr-qc/1611.07367]}}.
\newblock
  doi:{\changeurlcolor{black}\href{https://doi.org/10.1103/PhysRevD.95.044045}{\detokenize{10.1103/PhysRevD.95.044045}}}.

\bibitem[Hannam \em{et~al.}(2014)Hannam, Schmidt, Bohé, Haegel, Husa, Ohme,
  Pratten, and Pürrer]{Hannam:2013oca}
Hannam, M.; Schmidt, P.; Bohé, A.; Haegel, L.; Husa, S.; Ohme, F.; Pratten,
  G.; Pürrer, M.
\newblock {Simple Model of Complete Precessing Black-Hole-Binary Gravitational
  Waveforms}.
\newblock {\em Phys. Rev. Lett.} {\bf 2014}, {\em 113},~151101,
  \href{http://xxx.lanl.gov/abs/1308.3271}{{\normalfont
  [arXiv:gr-qc/1308.3271]}}.
\newblock
  doi:{\changeurlcolor{black}\href{https://doi.org/10.1103/PhysRevLett.113.151101}{\detokenize{10.1103/PhysRevLett.113.151101}}}.

\bibitem[Abbott \em{et~al.}(2018)Abbott et~al.]{Abbott:2018lct}
Abbott, B.P.; others.
\newblock {Tests of General Relativity with GW170817} {\bf 2018}.
\newblock  \href{http://xxx.lanl.gov/abs/1811.00364}{{\normalfont
  [arXiv:gr-qc/1811.00364]}}.

\bibitem[Takami \em{et~al.}(2014)Takami, Rezzolla, and Baiotti]{Takami:2014zpa}
Takami, K.; Rezzolla, L.; Baiotti, L.
\newblock {Constraining the Equation of State of Neutron Stars from Binary
  Mergers}.
\newblock {\em Phys. Rev. Lett.} {\bf 2014}, {\em 113},~091104,
  \href{http://xxx.lanl.gov/abs/1403.5672}{{\normalfont
  [arXiv:gr-qc/1403.5672]}}.
\newblock
  doi:{\changeurlcolor{black}\href{https://doi.org/10.1103/PhysRevLett.113.091104}{\detokenize{10.1103/PhysRevLett.113.091104}}}.

\bibitem[Takami \em{et~al.}(2015)Takami, Rezzolla, and Baiotti]{Takami:2014tva}
Takami, K.; Rezzolla, L.; Baiotti, L.
\newblock {Spectral properties of the post-merger gravitational-wave signal
  from binary neutron stars}.
\newblock {\em Phys. Rev.} {\bf 2015}, {\em D91},~064001,
  \href{http://xxx.lanl.gov/abs/1412.3240}{{\normalfont
  [arXiv:gr-qc/1412.3240]}}.
\newblock
  doi:{\changeurlcolor{black}\href{https://doi.org/10.1103/PhysRevD.91.064001}{\detokenize{10.1103/PhysRevD.91.064001}}}.

\bibitem[Rezzolla and Takami(2016)]{Rezzolla:2016nxn}
Rezzolla, L.; Takami, K.
\newblock {Gravitational-wave signal from binary neutron stars: a systematic
  analysis of the spectral properties}.
\newblock {\em Phys. Rev.} {\bf 2016}, {\em D93},~124051,
  \href{http://xxx.lanl.gov/abs/1604.00246}{{\normalfont
  [arXiv:gr-qc/1604.00246]}}.
\newblock
  doi:{\changeurlcolor{black}\href{https://doi.org/10.1103/PhysRevD.93.124051}{\detokenize{10.1103/PhysRevD.93.124051}}}.

\bibitem[Bauswein and Janka(2012)]{Bauswein:2011tp}
Bauswein, A.; Janka, H.T.
\newblock {Measuring neutron-star properties via gravitational waves from
  binary mergers}.
\newblock {\em Phys. Rev. Lett.} {\bf 2012}, {\em 108},~011101,
  \href{http://xxx.lanl.gov/abs/1106.1616}{{\normalfont
  [arXiv:astro-ph.SR/1106.1616]}}.
\newblock
  doi:{\changeurlcolor{black}\href{https://doi.org/10.1103/PhysRevLett.108.011101}{\detokenize{10.1103/PhysRevLett.108.011101}}}.

\bibitem[Stergioulas \em{et~al.}(2011)Stergioulas, Bauswein, Zagkouris, and
  Janka]{Stergioulas:2011gd}
Stergioulas, N.; Bauswein, A.; Zagkouris, K.; Janka, H.T.
\newblock {Gravitational waves and nonaxisymmetric oscillation modes in mergers
  of compact object binaries}.
\newblock {\em Mon. Not. Roy. Astron. Soc.} {\bf 2011}, {\em 418},~427,
  \href{http://xxx.lanl.gov/abs/1105.0368}{{\normalfont
  [arXiv:gr-qc/1105.0368]}}.
\newblock
  doi:{\changeurlcolor{black}\href{https://doi.org/10.1111/j.1365-2966.2011.19493.x}{\detokenize{10.1111/j.1365-2966.2011.19493.x}}}.

\bibitem[Bauswein \em{et~al.}(2012)Bauswein, Janka, Hebeler, and
  Schwenk]{Bauswein:2012ya}
Bauswein, A.; Janka, H.T.; Hebeler, K.; Schwenk, A.
\newblock {Equation-of-state dependence of the gravitational-wave signal from
  the ring-down phase of neutron-star mergers}.
\newblock {\em Phys. Rev.} {\bf 2012}, {\em D86},~063001,
  \href{http://xxx.lanl.gov/abs/1204.1888}{{\normalfont
  [arXiv:astro-ph.SR/1204.1888]}}.
\newblock
  doi:{\changeurlcolor{black}\href{https://doi.org/10.1103/PhysRevD.86.063001}{\detokenize{10.1103/PhysRevD.86.063001}}}.

\bibitem[Bauswein \em{et~al.}(2014)Bauswein, Stergioulas, and
  Janka]{Bauswein:2014qla}
Bauswein, A.; Stergioulas, N.; Janka, H.T.
\newblock {Revealing the high-density equation of state through binary neutron
  star mergers}.
\newblock {\em Phys. Rev.} {\bf 2014}, {\em D90},~023002,
  \href{http://xxx.lanl.gov/abs/1403.5301}{{\normalfont
  [arXiv:astro-ph.SR/1403.5301]}}.
\newblock
  doi:{\changeurlcolor{black}\href{https://doi.org/10.1103/PhysRevD.90.023002}{\detokenize{10.1103/PhysRevD.90.023002}}}.

\bibitem[Clark \em{et~al.}(2016)Clark, Bauswein, Stergioulas, and
  Shoemaker]{Clark:2015zxa}
Clark, J.A.; Bauswein, A.; Stergioulas, N.; Shoemaker, D.
\newblock {Observing Gravitational Waves From The Post-Merger Phase Of Binary
  Neutron Star Coalescence}.
\newblock {\em Class. Quant. Grav.} {\bf 2016}, {\em 33},~085003,
  \href{http://xxx.lanl.gov/abs/1509.08522}{{\normalfont
  [arXiv:astro-ph.HE/1509.08522]}}.
\newblock
  doi:{\changeurlcolor{black}\href{https://doi.org/10.1088/0264-9381/33/8/085003}{\detokenize{10.1088/0264-9381/33/8/085003}}}.

\bibitem[Bauswein and Stergioulas(2015)]{Bauswein:2015yca}
Bauswein, A.; Stergioulas, N.
\newblock {Unified picture of the post-merger dynamics and gravitational wave
  emission in neutron star mergers}.
\newblock {\em Phys. Rev.} {\bf 2015}, {\em D91},~124056,
  \href{http://xxx.lanl.gov/abs/1502.03176}{{\normalfont
  [arXiv:astro-ph.SR/1502.03176]}}.
\newblock
  doi:{\changeurlcolor{black}\href{https://doi.org/10.1103/PhysRevD.91.124056}{\detokenize{10.1103/PhysRevD.91.124056}}}.

\bibitem[Bernuzzi \em{et~al.}(2015)Bernuzzi, Dietrich, and
  Nagar]{Bernuzzi:2015rla}
Bernuzzi, S.; Dietrich, T.; Nagar, A.
\newblock {Modeling the complete gravitational wave spectrum of neutron star
  mergers}.
\newblock {\em Phys. Rev. Lett.} {\bf 2015}, {\em 115},~091101,
  \href{http://xxx.lanl.gov/abs/1504.01764}{{\normalfont
  [arXiv:gr-qc/1504.01764]}}.
\newblock
  doi:{\changeurlcolor{black}\href{https://doi.org/10.1103/PhysRevLett.115.091101}{\detokenize{10.1103/PhysRevLett.115.091101}}}.

\bibitem[Chaurasia \em{et~al.}(2018)Chaurasia, Dietrich, Johnson-McDaniel,
  Ujevic, Tichy, and Brügmann]{Chaurasia:2018zhg}
Chaurasia, S.V.; Dietrich, T.; Johnson-McDaniel, N.K.; Ujevic, M.; Tichy, W.;
  Brügmann, B.
\newblock {Gravitational waves and mass ejecta from binary neutron star
  mergers: Effect of large eccentricities}.
\newblock {\em Phys. Rev.} {\bf 2018}, {\em D98},~104005,
  \href{http://xxx.lanl.gov/abs/1807.06857}{{\normalfont
  [arXiv:gr-qc/1807.06857]}}.
\newblock
  doi:{\changeurlcolor{black}\href{https://doi.org/10.1103/PhysRevD.98.104005}{\detokenize{10.1103/PhysRevD.98.104005}}}.

\bibitem[Metzger(2017)]{Metzger:2016pju}
Metzger, B.D.
\newblock {Kilonovae}.
\newblock {\em Living Rev. Rel.} {\bf 2017}, {\em 20},~3,
  \href{http://xxx.lanl.gov/abs/1610.09381}{{\normalfont
  [arXiv:astro-ph.HE/1610.09381]}}.
\newblock
  doi:{\changeurlcolor{black}\href{https://doi.org/10.1007/s41114-017-0006-z}{\detokenize{10.1007/s41114-017-0006-z}}}.

\bibitem[Dietrich and Ujevic(2017)]{Dietrich:2016fpt}
Dietrich, T.; Ujevic, M.
\newblock {Modeling dynamical ejecta from binary neutron star mergers and
  implications for electromagnetic counterparts}.
\newblock {\em Class. Quant. Grav.} {\bf 2017}, {\em 34},~105014,
  \href{http://xxx.lanl.gov/abs/1612.03665}{{\normalfont
  [arXiv:gr-qc/1612.03665]}}.
\newblock
  doi:{\changeurlcolor{black}\href{https://doi.org/10.1088/1361-6382/aa6bb0}{\detokenize{10.1088/1361-6382/aa6bb0}}}.

\bibitem[Kohri \em{et~al.}(2005)Kohri, Narayan, and Piran]{Kohri:2005tq}
Kohri, K.; Narayan, R.; Piran, T.
\newblock {Neutrino-dominated accretion and supernovae}.
\newblock {\em Astrophys. J.} {\bf 2005}, {\em 629},~341--361,
  \href{http://xxx.lanl.gov/abs/astro-ph/0502470}{{\normalfont
  [arXiv:astro-ph/astro-ph/0502470]}}.
\newblock
  doi:{\changeurlcolor{black}\href{https://doi.org/10.1086/431354}{\detokenize{10.1086/431354}}}.

\bibitem[Surman \em{et~al.}(2006)Surman, McLaughlin, and Hix]{Surman:2005kf}
Surman, R.; McLaughlin, G.C.; Hix, W.R.
\newblock {Nucleosynthesis in the outflow from gamma-ray burst accretion
  disks}.
\newblock {\em Astrophys. J.} {\bf 2006}, {\em 643},~1057--1064,
  \href{http://xxx.lanl.gov/abs/astro-ph/0509365}{{\normalfont
  [arXiv:astro-ph/astro-ph/0509365]}}.
\newblock
  doi:{\changeurlcolor{black}\href{https://doi.org/10.1086/501116}{\detokenize{10.1086/501116}}}.

\bibitem[Metzger \em{et~al.}(2008)Metzger, Piro, and Quataert]{Metzger:2008av}
Metzger, B.D.; Piro, A.L.; Quataert, E.
\newblock {Time-Dependent Models of Accretion Disks Formed from Compact Object
  Mergers}.
\newblock {\em Mon. Not. Roy. Astron. Soc.} {\bf 2008}, {\em 390},~781,
  \href{http://xxx.lanl.gov/abs/0805.4415}{{\normalfont
  [arXiv:astro-ph/0805.4415]}}.
\newblock
  doi:{\changeurlcolor{black}\href{https://doi.org/10.1111/j.1365-2966.2008.13789.x}{\detokenize{10.1111/j.1365-2966.2008.13789.x}}}.

\bibitem[Dessart \em{et~al.}(2009)Dessart, Ott, Burrows, Rosswog, and
  Livne]{Dessart:2008zd}
Dessart, L.; Ott, C.D.; Burrows, A.; Rosswog, S.; Livne, E.
\newblock {Neutrino signatures and the neutrino-driven wind in Binary Neutron
  Star Mergers}.
\newblock {\em Astrophys. J.} {\bf 2009}, {\em 690},~1681,
  \href{http://xxx.lanl.gov/abs/0806.4380}{{\normalfont
  [arXiv:astro-ph/0806.4380]}}.
\newblock
  doi:{\changeurlcolor{black}\href{https://doi.org/10.1088/0004-637X/690/2/1681}{\detokenize{10.1088/0004-637X/690/2/1681}}}.

\bibitem[Fernández and Metzger(2013)]{Fernandez:2013tya}
Fernández, R.; Metzger, B.D.
\newblock {Delayed outflows from black hole accretion tori following neutron
  star binary coalescence}.
\newblock {\em Mon. Not. Roy. Astron. Soc.} {\bf 2013}, {\em 435},~502,
  \href{http://xxx.lanl.gov/abs/1304.6720}{{\normalfont
  [arXiv:astro-ph.HE/1304.6720]}}.
\newblock
  doi:{\changeurlcolor{black}\href{https://doi.org/10.1093/mnras/stt1312}{\detokenize{10.1093/mnras/stt1312}}}.

\bibitem[Perego \em{et~al.}(2014)Perego, Rosswog, Cabezón, Korobkin, Käppeli,
  Arcones, and Liebendörfer]{Perego:2014fma}
Perego, A.; Rosswog, S.; Cabezón, R.M.; Korobkin, O.; Käppeli, R.; Arcones,
  A.; Liebendörfer, M.
\newblock {Neutrino-driven winds from neutron star merger remnants}.
\newblock {\em Mon. Not. Roy. Astron. Soc.} {\bf 2014}, {\em 443},~3134--3156,
  \href{http://xxx.lanl.gov/abs/1405.6730}{{\normalfont
  [arXiv:astro-ph.HE/1405.6730]}}.
\newblock
  doi:{\changeurlcolor{black}\href{https://doi.org/10.1093/mnras/stu1352}{\detokenize{10.1093/mnras/stu1352}}}.

\bibitem[Siegel \em{et~al.}(2014)Siegel, Ciolfi, and Rezzolla]{Siegel:2014ita}
Siegel, D.M.; Ciolfi, R.; Rezzolla, L.
\newblock {Magnetically driven winds from differentially rotating neutron stars
  and X-ray afterglows of short gamma-ray bursts}.
\newblock {\em Astrophys. J.} {\bf 2014}, {\em 785},~L6,
  \href{http://xxx.lanl.gov/abs/1401.4544}{{\normalfont
  [arXiv:astro-ph.HE/1401.4544]}}.
\newblock
  doi:{\changeurlcolor{black}\href{https://doi.org/10.1088/2041-8205/785/1/L6}{\detokenize{10.1088/2041-8205/785/1/L6}}}.

\bibitem[Just \em{et~al.}(2015)Just, Bauswein, Pulpillo, Goriely, and
  Janka]{Just:2014fka}
Just, O.; Bauswein, A.; Pulpillo, R.A.; Goriely, S.; Janka, H.T.
\newblock {Comprehensive nucleosynthesis analysis for ejecta of compact binary
  mergers}.
\newblock {\em Mon. Not. Roy. Astron. Soc.} {\bf 2015}, {\em 448},~541--567,
  \href{http://xxx.lanl.gov/abs/1406.2687}{{\normalfont
  [arXiv:astro-ph.SR/1406.2687]}}.
\newblock
  doi:{\changeurlcolor{black}\href{https://doi.org/10.1093/mnras/stv009}{\detokenize{10.1093/mnras/stv009}}}.

\bibitem[Rezzolla and Kumar(2015)]{Rezzolla:2014nva}
Rezzolla, L.; Kumar, P.
\newblock {A novel paradigm for short gamma-ray bursts with extended X-ray
  emission}.
\newblock {\em Astrophys. J.} {\bf 2015}, {\em 802},~95,
  \href{http://xxx.lanl.gov/abs/1410.8560}{{\normalfont
  [arXiv:astro-ph.HE/1410.8560]}}.
\newblock
  doi:{\changeurlcolor{black}\href{https://doi.org/10.1088/0004-637X/802/2/95}{\detokenize{10.1088/0004-637X/802/2/95}}}.

\bibitem[Ciolfi and Siegel(2015)]{Ciolfi:2014yla}
Ciolfi, R.; Siegel, D.M.
\newblock {Short gamma-ray bursts in the "time-reversal" scenario}.
\newblock {\em Astrophys. J.} {\bf 2015}, {\em 798},~L36,
  \href{http://xxx.lanl.gov/abs/1411.2015}{{\normalfont
  [arXiv:astro-ph.HE/1411.2015]}}.
\newblock
  doi:{\changeurlcolor{black}\href{https://doi.org/10.1088/2041-8205/798/2/L36}{\detokenize{10.1088/2041-8205/798/2/L36}}}.

\bibitem[Siegel and Metzger(2017)]{Siegel:2017nub}
Siegel, D.M.; Metzger, B.D.
\newblock {Three-Dimensional General-Relativistic Magnetohydrodynamic
  Simulations of Remnant Accretion Disks from Neutron Star Mergers: Outflows
  and $r$-Process Nucleosynthesis}.
\newblock {\em Phys. Rev. Lett.} {\bf 2017}, {\em 119},~231102,
  \href{http://xxx.lanl.gov/abs/1705.05473}{{\normalfont
  [arXiv:astro-ph.HE/1705.05473]}}.
\newblock
  doi:{\changeurlcolor{black}\href{https://doi.org/10.1103/PhysRevLett.119.231102}{\detokenize{10.1103/PhysRevLett.119.231102}}}.

\bibitem[Kasen \em{et~al.}(2017)Kasen, Metzger, Barnes, Quataert, and
  Ramirez-Ruiz]{Kasen:2017sxr}
Kasen, D.; Metzger, B.; Barnes, J.; Quataert, E.; Ramirez-Ruiz, E.
\newblock {Origin of the heavy elements in binary neutron-star mergers from a
  gravitational wave event}.
\newblock {\em Nature} {\bf 2017},
  \href{http://xxx.lanl.gov/abs/1710.05463}{{\normalfont
  [arXiv:astro-ph.HE/1710.05463]}}.
\newblock [Nature551,80(2017)],
  doi:{\changeurlcolor{black}\href{https://doi.org/10.1038/nature24453}{\detokenize{10.1038/nature24453}}}.

\bibitem[Paczynski(1986)]{Paczynski:1986px}
Paczynski, B.
\newblock {Gamma-ray bursters at cosmological distances}.
\newblock {\em Astrophys. J.} {\bf 1986}, {\em 308},~L43--L46.
\newblock
  doi:{\changeurlcolor{black}\href{https://doi.org/10.1086/184740}{\detokenize{10.1086/184740}}}.

\bibitem[Eichler \em{et~al.}(1989)Eichler, Livio, Piran, and
  Schramm]{Eichler:1989ve}
Eichler, D.; Livio, M.; Piran, T.; Schramm, D.N.
\newblock {Nucleosynthesis, Neutrino Bursts and Gamma-Rays from Coalescing
  Neutron Stars}.
\newblock {\em Nature} {\bf 1989}, {\em 340},~126--128.
\newblock [,682(1989)],
  doi:{\changeurlcolor{black}\href{https://doi.org/10.1038/340126a0}{\detokenize{10.1038/340126a0}}}.

\bibitem[van Eerten \em{et~al.}(2018)van Eerten, Ryan, Ricci, Burgess,
  Wieringa, Piro, Cenko, and Sakamoto]{vanEerten:2018vgj}
van Eerten, E.T.H.; Ryan, G.; Ricci, R.; Burgess, J.M.; Wieringa, M.; Piro, L.;
  Cenko, S.B.; Sakamoto, T.
\newblock {A year in the life of GW170817: the rise and fall of a structured
  jet from a binary neutron star merger} {\bf 2018}.
\newblock  \href{http://xxx.lanl.gov/abs/1808.06617}{{\normalfont
  [arXiv:astro-ph.HE/1808.06617]}}.

\end{thebibliography}

\authorcontributions{Conceptualization, T.D. and M.U.; 
methodology, T.D.; 
investigation, H.G., T.D., M.U.; 
writing--original draft preparation, H.G, T.D., M.U.; 
writing--review and editing, H.G, T.D., M.U.; 
visualization, H.G, T.D.; 
funding acquisition, T.D., M.U.}

\funding{This research was funded by the European Union's Horizon 
  2020 research and innovation program under grant
  agreement No 749145, BNSmergers.
  Computations were performed on 
  the supercomputer SuperMUC at the LRZ
  (Munich) under the project number pr48pu. M.U. and H.G. also thanks 
CAPES 
for financial support.
  }

\acknowledgments{We thank Michael Coughlin for providing us with the lightcurve data of Fig.~8. 
We acknowledgment fruitful discussions with Sebastiano Bernuzzi, Bernd Br\"ugmann, Wolfgang Tichy, 
and Tsun Ho Pang.}

\conflictsofinterest{The authors declare no conflict of interest.}


\abbreviations{The following abbreviations are used in this manuscript:\\

\noindent
\begin{tabular}{@{}ll}
BH & Black Hole \\
BNS & Binary Neutron Star \\
CFL & Color-flavor-locked \\
EM & Electromagnetic \\
EoS & Equation of State \\
GRB & Gamma Ray Burst \\
GRHD & General Relativistic Hydrodynamics \\
GW & Gravitational Wave \\
HM & Hadronic Matter \\
HRSC & High-resolution Shock-capturing \\
HyS & Hybrid Star \\
LLF & Local Lax-Friedrich \\
MIT & Massachusetts Institute of Technology \\
NR & Numerical Relativity \\
NS & Neutron Star \\
QCD & Quantum Chromodynamics \\
SQM & Strange Quark Matter \\
WENOZ & Weighted Essentially Non-oscillatory \\
\end{tabular}}

\appendix
\section{HyS Piecewise Polytrope}\label{appA}
In the following, we present the EOS parameters defining the employed HyS EOS; 
cf.~Eq.~\eqref{eq:pwp} in section (\ref{subsec:HyS pwp}).
\begin{table}[h]
\caption{Piecewise polytrope of the HyS EoS used in the simulations. The units are chosen so that the rest-mass density $\rho$ is in $\rm g/cm^3$, $\Gamma$ is dimensionless and $K$ is such that the pressure $p$ is in $\rm g/cm^3$. The $i$-th row refers to the polytrope $p = K_i \rho^{\Gamma_i}$, for $\rho_i \leq \rho \leq \rho_{i+1}$.}
\label{tab:pwp}
\centering
\begin{tabular}{c|c|c}
\hline 
$\rho$ & $K$ & $\Gamma$ \\
\hline
$0.00000$ & $3.99873\times10^{-8}$ & $1.357$ \\
$1.46231\times10^{14}$ & $1.80658\times10^{-31}$ & $3.005$ \\
$2.60400\times10^{14}$ & $3.76582\times10^{12}$ & $0.000$ \\
$4.66533\times10^{14}$ & $1.00162\times10^{-167}$ & $12.242$ \\
$5.06058\times10^{14}$ & $1.29766\times10^{-108}$ & $8.222$ \\
$6.07892\times10^{14}$ & $2.18919\times10^{-43}$ & $3.810$ \\
$7.22396\times10^{14}$ & $7.07726\times10^{-28}$ & $2.766$ \\
$8.50306\times10^{14}$ & $7.86344\times10^{-21}$ & $2.294$ \\
$9.92358\times10^{14}$ & $7.59597\times10^{-17}$ & $2.028$ \\
$1.14929\times10^{15}$ & $2.58041\times10^{-14}$ & $1.860$ \\
$1.32183\times10^{15}$ & $1.38093\times10^{-12}$ & $1.746$ \\
$1.51072\times10^{15}$ & $2.40271\times10^{-11}$ & $1.664$ \\
$1.71669\times10^{15}$ & $2.01349\times10^{-10}$ & $1.604$ \\
$1.94048\times10^{15}$ & $1.02422\times10^{-9}$ & $1.557$ \\
$2.18282\times10^{15}$ & $4.06158\times10^{-9}$ & $1.518$ \\
\hline 
\end{tabular}
\end{table}
\end{document}